\documentclass[aps,prb,groupedaddress,twocolumn]{revtex4-1}
\pdfoutput=1

\usepackage{amsmath}
\usepackage{amssymb}
\usepackage{graphicx}
\usepackage{bm}
\usepackage{hyperref}
\hypersetup{%
breaklinks = {true},
citecolor = {blue},
colorlinks = {true},
linkcolor = {red},
pdfauthor = {\textcopyright\ Francesco Catalano},
pdfcreator = {\LaTeX\ and \flqq hyperref\frqq},
pdffitwindow = {true},
pdfmenubar = {true},
pdfpagelayout = {SinglePage},
pdfstartview = {Fit},
pdftoolbar = {true},
plainpages = {false},
}

\usepackage{bbm}

\newcommand{\la}{\langle}
\newcommand{\ra}{\rangle}

\newcommand{\dla}{\langle \negthinspace \langle}
\newcommand{\dra}{\rangle \negthinspace \rangle}

\newcommand{\ua}{\uparrow}
\newcommand{\da}{\downarrow}

\newcommand{\vk}{{\bf k}}
\newcommand{\vp}{{\bf p}}
\newcommand{\vx}{{\bf x}}

\usepackage{bbold}
\newcommand{\eye}{\mathbb{1}}

\usepackage{color}
\usepackage{color}

\begin{document}


\title{Equation of motion truncation scheme based on partial orthogonalization}

\author{Francesco Catalano}
\affiliation{Department of Physics and Astronomy, Uppsala University,
P.O. Box 516, SE-75120, Uppsala, Sweden}
\author{Johan Nilsson}
\affiliation{Department of Physics and Astronomy, Uppsala University,
P.O. Box 516, SE-75120, Uppsala, Sweden}

\date{March 9, 2020}

\begin{abstract}

We introduce a general scheme to consistently truncate equations of motion for Green's functions.
Our scheme is guaranteed to generate physical Green's functions with real excitation energies and positive spectral weights. There are free parameters in our scheme akin to mean field parameters that may be determined to get as good an approximation to the physics as possible.
As a test case we apply our scheme to a two-pole approximation for the 2D Hubbard model.
At half-filling we find an insulating solution with several interesting properties: it has low expectation value of the energy and it gives upper and lower Hubbard bands with the full non-interacting bandwidth in the large $U$ limit. Away from half-filling, in particular in the intermediate interaction regime, our scheme allows for several different phases with different number of Fermi surfaces and topologies.

\end{abstract}

\maketitle

%
%
\section{Introduction}

Green's function methods are widely used to study many-body systems and they represent a natural framework that connects microscopical details of a theory with its macroscopical properties.\cite{Martin1959}

The attempt to self-consistently determine these quantities has a long history and it is still remains one of the central paradigms in the study of strongly correlated systems. From the most recent DMFT,\cite{DMFT_RMP}  where they are used to fix the mapping of a lattice model onto an impurity one, to the older equation of motion approach.\cite{Tyablikov1959,Tyablikov1967}
In the latter method, given an interacting Hamiltonian, an extensively growing chain of coupled equations are derived.\cite{Tserkovnikov1981,Zubarev}
For few-body systems it is possible to use various implementations of this method to obtain the single particle Green's function exactly.\cite{dimerEOM}

However, in order to study thermodynamical properties of an interacting system a truncation procedure able to approximately decouple this extensively growing system of coupled equations plays a crucial role.
Early attempts in the construction of truncation schemes explored arbitrary truncation schemes and decoupling schemes of Tyablikov-type.\cite{Tyablikov1959,Tyablikov1967}
Despite some successful applications these decoupling schemes often led to violation of the analytical structure of the Green's functions, predicting imaginary poles and negative spectral weight for the single particle Green's function.
Despite these difficulties Hubbard in his pioneering  work,\cite{Hubbard_I}
managed to find a useful decoupling for a two-pole approximation for the Hubbard model.
This decoupling (Hubbard-I) is still often used in treating strongly correlation in presence of local interactions, especially in studies of quantum systems out of equilibrium,\cite{Fransson2019} and multi-orbital systems.\cite{rare_earth_hubbardI}

Almost a decade after these early works Roth developed a universal decoupling scheme able to enforce correct analytical properties for approximated Green's functions.\cite{PhysRevLett.20.1431} This decoupling scheme is now called the Roth procedure, and often relies on parameters that can not be determined within the scheme itself, making unavoidable ulterior approximations. For this reason this method is often regarded as an uncontrolled approximation, which severely limits its applicability.
The works of  Mancini and Avella {\it et al.}\cite{Mancini_Avella_1}  show that the Roth procedure leads to violations of other physical principles such as the Pauli principle and that it is possible to constrain some, if not all of the unknown decoupling parameters, by enforcing such physical requirements.
Despite much progress in finding easy extendable decoupling schemes,\cite{Fan2018} the possibility to systematically check what are the approximations involved in the decoupling still remains a neglected aspect.

In this paper we present a decoupling scheme based on a partial orthogonalization of the operators involved, where the relation between the true Green's function and the approximate one can readily be obtained.
The paper is organized as follows:
in Sec.~\ref{secEOM} we provide a general discussion of the formalism, we clarify the role of the Hermiticity of the $E$-matrix and we present our decoupling scheme based on the partial orthogonalization of the operators.
In Sec.~\ref{sec:Hub2pole} we apply our scheme to a two-pole approximation of the Hubbard model making evident the relationship between the approximate and the true Green's function.
In Sec.~\ref{sec:sum_rules} we analyze the global sum rules that should be respected in the two-pole approximation of the Hubbard model and we present a variational scheme as a guiding principle for the determination of the unknown orthogonalization parameters.
In Sec.~\ref{half_fill_num} we provide numerical results at half-filling and in Sec.~\ref{sec:analytical} we give analytical formulas that are useful to understand the Green's function.
In Sec.~\ref{sec:holedoped_12} and Sec.~\ref{sec:holedoped_4} we discuss numerical results for hole doping in the strong- and intermediate-coupling regimes respectively.
Finally, in Sec.~\ref{sec:conclusions} we provide some conclusions and an outlook.

\section{A scheme for the truncation of the EoM}
\label{secEOM}

\subsection{Formalism review}
\label{sec:formalism}
We will mainly use the notation of Tserkovnikov in the following.\cite{Tserkovnikov1981}
For completeness we briefly review what we will need for this paper.
Let us first assume we have a set of fermionic operators  $\{\hat{A}_i\}_{i=1}^M$ closed under the commutation with the hamiltonian for some evolution matrix $K$
\begin{equation}
[\hat{A}_i,H]=\sum_j K_{ij}\hat{A}_j .
\end{equation}
Then the equation of motion (EOM) for the Green's function matrix gives
\begin{equation} 
z \dla  A^{\,}_i | A_j^\dagger \dra_z = \langle A^{\,}_i | A_j^\dagger \rangle   
+ \sum_k K_{ik} \dla A^{\,}_k | A_j^\dagger  \dra_z ,
\end{equation}
here the normalization matrix $N$ 
\begin{equation}
N_{ij} = \langle A^{\,}_i | A^\dagger_j \rangle
=
\langle  \{ \hat{A}^{\,}_i ,\hat{A}_j^\dagger \} \rangle .
\end{equation}
Consequently the Green's function, viewed as a matrix becomes
\begin{equation}
\dla A | A^\dagger \dra_z = \frac{1}{z\eye - K} N
=
N \frac{1}{z\eye - K^\dagger} ,
\end{equation}
where the second form is obtained making use of the fact that $\hat{H}$ is Hermitean.
For these two forms to be consistent we have the condition that
\begin{equation}
\label{phys-cond}
KN=NK^\dagger ,
\end{equation}
which will be of crucial importance in the developments below.
Finally one may calculate averages of bilinear of all of the operators involved using the formula
\begin{equation}
\label{form:eq2} 
\langle \hat{A}_{j}^{\dagger} \hat{A}_i^{\,} \rangle = \frac{1}{2\pi i}\oint dz f(z)  \dla A^{\,}_i | A_{j}^{\dagger}\dra_z ,
\end{equation}
where the contour encircles the real axis.

%
%

\subsection{A partial orthogonalization scheme for the truncation of the EOM}
As shown in the previous work this framework gives exact results if the set  of operator  $\{\hat{A}_i\}_{i=1}^M$  are closed under the commutation with the Hamiltonian.\cite{dimerEOM}
In an extend many-body system the number of operators necessary to close the equation of motion exactly will typically grow exponentially with the size of the system, making a direct application of this scheme unfeasible.

To produce a truncation scheme capable of producing physical Green's functions, it is important to notice that in a Hermitean theory the average of the operators involved in the dynamics  and their evolution are not independent. In particular as noticed by Roth \cite{PhysRevLett.20.1431} the matrix 
\begin{equation}
E_{ij} \equiv \langle \{ [\hat{A}^{\,}_i,\hat{H}],\hat{A}_j^\dagger \} \rangle = \sum_k K_{ik}N_{kj}
\end{equation}
needs to be Hermitean. Here $\langle \dots \rangle $ indicate some average over exact eigenstates of the theory $\hat{H}$. $K$ is the full evolution matrix of the operators and $N$ is the normalization matrix introduced above.
Using the fact that the matrix $N$ is Hermitean by construction (i.e., it holds for averages in any state) this gives the same consistency condition as Eq.~\eqref{phys-cond} above.
This condition together with the fact that $N$ has to be positive definite guarantees that the Green's function posseses real poles and positive spectral weight.\cite{PhysRevLett.20.1431}

When the hierarchy of the evolution of an operator $\hat{A}_1$ is considered at most one new operator is generated in each step, i.e., 
\begin{subequations}
\begin{eqnarray}
\bigl[ \hat{A}_1 ,\hat{H} \bigr] &=& K_{11} \hat{A}_1 + K_{12} \hat{A}_2 , \\
\bigl[ \hat{A}_2 , \hat{H} \bigr] &=& K_{21} \hat{A}_1 + K_{22} \hat{A}_2 + K_{23} \hat{A}_3 ,
\end{eqnarray}
\end{subequations}
etc. until the EOM closes and no new operators are generated. Note that $\hat{A}_2$ is not unique since one can add a part of $\hat{A}_1$ to it, and similarly for the other higher $\hat{A}$'s. In any event $K$ is only non-zero on the first upper diagonal and below.
Let us now truncate the EOM at the $q$-th operator. A brute force truncation of the matrices involved gives
\begin{equation}
 K_{trunc}=\begin{pmatrix}
 K_{11} & K_{12}&  &0\\
 K_{21}& K_{22} & &0\\
 \vdots&  &\ddots& \vdots \\
K_{q1}&K_{q2} &\hdots&K_{qq}
\end{pmatrix},
\end{equation}
and the corresponding $N_{trunc}$
\begin{equation}
 N_{trunc}=\begin{pmatrix}N_{11}&N_{12}&  &N_{1q}\\
 N_{21}&N_{22} & &N_{2q}\\
 \vdots&  &\ddots& \vdots \\
N_{q1}&N_{q2} &\hdots&N_{qq} \end{pmatrix}.
\end{equation}
Now we note that
\begin{equation}
E_{trunc}=K_{trunc}N_{trunc},
\end{equation}  
differs from the corresponding sub-block of the full $E$ only in the last row, through the coupling of $K_{q , q+1}$ to the $(q+1)$-th column of the full $N$ matrix.
Therefore an arbitrary truncation of the equation of motion is going to generate an evolution that in general does not satisfy the condition in Eq.~\eqref{phys-cond}, leading to a potentially unphysical approximation for the Green's function.

In this paper we propose to restore the Hermiticity of $E_{trunc}$ adding to the first operator not considered explicitly in the dynamics $\hat{A}_{q+1}$ a linear combination of the operators $\hat{A}_1,\dots,\hat{A}_q$
\begin{equation}
\hat{A}'_{q+1} = \hat{A}_{q+1} -\sum_{l=1}^q \lambda_l \hat{A}_l .
\end{equation}
Most of the $\lambda$ parameters will be fixed by demanding that
\begin{equation}
\langle A'_{q+1} |A^\dagger_j\rangle =0 ~~~~~~~~\text{for} ~~~j=1,\dots,q-1.
\label{eq:partial_ortho1}
\end{equation}
This partial orthogonalization procedure ensures that $E_{trunc}$ is Hermitean, because it makes it identical to the corresponding block of $E$ except for the last element on the diagonal $E_{qq}$ which is not fixed by our procedure.
The Roth procedure corresponds to also orthogonalizing with respect to $\hat{A}_q^\dagger$. This gives $q$ equations for $q$ unknowns, and therefore also fixes the value of $E_{qq}$, whereas in our scheme we have $q-1$ equations for $q$ unknowns, leaving $E_{qq}$ arbitrary. We will use this additional freedom to make sure that our approximation fulfills other physically relevant criteria such as Pauli principle constraints or sum rules.

In the next section we are going to elucidate this procedure by applying it to a two-pole approximation of the Hubbard model. In particular it will be evident that the effect of this procedure is a non-unique modification of the last row of $K_{trunc}$. This arbitrariness can be exploited to enforce global sum rules for the Green's functions and open up the possibility of using different criteria to fix the free parameters $\lambda_i$ not fixed by Eq.~\eqref{eq:partial_ortho1}.

A last remark on this scheme is that despite the freedom in the choice of the parameters $\lambda_i$ one can always write the residual Green's functions not considered explicitly in the dynamics, making transparent the approximation involved in this truncation of the equation of motion. 

%
%
\section{Application to the Hubbard model in a two-pole approximation}
\label{sec:Hub2pole}

In this section we will apply our scheme to a two-pole approximation to the Green's function in the Hubbard model.
Let us consider the Hubbard hamiltonian
\begin{equation}
\hat{H}= \sum_{\vk} \epsilon_\vk (c^\dagger_{\vk\uparrow}c^{\,}_{\vk\uparrow} + 
c^\dagger_{\vk\downarrow}c^{\,}_{\vk\downarrow} )+ U \sum_i n_{i\uparrow }n_{i\downarrow},
\end{equation}
We will denote the total number of sites with $N_s$, $c^{\,}_{\vk \sigma}$ indicates fermion operator with spin $\sigma$ and $n_{i\sigma} = c^\dagger_{i\sigma}c^{\,}_{i\sigma}$.\\
Let us consider the first three operators that appear in the equation of motion hierarchy 
\begin{eqnarray*}
\hat{A}_{1\vk} &=& c^{\,}_{\vk \uparrow},\\
\hat{A}_{2\vk} &=& (c_{\downarrow}^\dagger c^{\,}_{\downarrow}c^{\,}_{\uparrow})_{\vk}.\\
\hat{A}_{3 \vk} &=&\frac{1}{\sqrt{N_s}}\sum_\vp
\Bigr(  \epsilon_\vp 
\big[(c^{\dagger}_{\downarrow}c_{\downarrow})_{\vk-\vp}c_{\vp\uparrow}
- (c^{\dagger}_{\downarrow}c_{\uparrow})_{\vk-\vp}c_{\vp\downarrow}\bigr] 
\nonumber \\
 & &- \epsilon_{-\vp}c^{\dagger}_{-\vp\downarrow}(c_{\downarrow}c_{\uparrow})_{\vk-\vp} \Bigr). 
\end{eqnarray*}
where we have introduced
\begin{equation}
(\hat{O}_1 \hdots  \hat{O}_n )_\vk= \frac{1}{\sqrt{N_s}} \sum_{i}  e^{i \vk \cdot \vx_i } \hat{O}_{1x_i}  \dots \hat{O}_{nx_i} . 
\end{equation}
Let us first do a brute force truncation of the evolution after two operators.
The truncated evolution becomes
\begin{equation}
K_{trunc} (\vk)=\begin{pmatrix} \epsilon_\vk&U\\0&U\end{pmatrix},
\end{equation} 
and the respective N matrix becomes 
\begin{equation}
N_{trunc} (\vk) = N = 
\begin{pmatrix} 1& \bar{n}_{\downarrow}\\ 
\bar{n}_{\downarrow}&\bar{n}_{\downarrow}
\end{pmatrix},
\end{equation}
with $\bar{n}_\da = \la n_{i\da} \ra$ which is independent of the site index $i$.
In this case $E_{trunc}=K_{trunc} N_{trunc}$  is not Hermitean (except in special cases such as $U=0$, $\epsilon_\vk = 0$ or $\bar{n}_\da = 0$) and this leads to an unphysical approximation for the Green's function for some range of the parameters.

Let us now apply our scheme to this particular problem, in this case we need to determine $\lambda_{1\vk},\lambda_{2\vk}$ such that
\begin{equation}
\langle A_{3\vk}| c^{\dagger}_\vk \rangle - \lambda_{1\vk} \langle A_{1\vk}| c^{\dagger}_\vk \rangle 
-\lambda_{2\vk} \langle A_{2\vk}| c^{\dagger}_\vk \rangle =0.
\end{equation}
Evaluating the anticommutator averages we obtain
\begin{equation}
\label{cond:orth}
\epsilon_\vk \bar{n}_{\downarrow} -\lambda_{1\vk} - \lambda_{2\vk} \bar{n}_{\downarrow}=0 .
\end{equation}
As already anticipated in Sec.~\ref{secEOM}, the values of  $\lambda_{1\vk}$ and $\lambda_{2\vk}$ are not uniquely determined by this procedure. Without any loss of generality let us eliminate $\lambda_{1\vk}$ writing
\begin{equation}
\lambda_{1\vk} = (\epsilon_\vk-\lambda_{2\vk}) \bar{n}_{\downarrow} 
\end{equation}
Using this we can write
\begin{equation}
\hat{A}_{3\vk} = \hat{A}'_{3\vk} + (\epsilon_\vk-\lambda_{2\vk}) \bar{n}_{\downarrow} \hat{A}_{1\vk} + \lambda_{2\vk} \hat{A}_{2\vk}
\end{equation}
where  $\langle A'_{3\vk}| c^\dagger_\vk \rangle=0$.\\
At this point the equation of motion for the operator $\hat{A}_{1\vk}$  can be rewritten as ($B$ is here arbitrary)
\begin{eqnarray}
z \dla A_{1\vk} | B^\dagger \dra &=& \la A_{1\vk} | B^\dagger \ra 
\nonumber \\
&+& \epsilon_\vk  \dla A_{1\vk} | B^\dagger \dra +
U  \dla A_{2 \vk} | B^\dagger \dra
\end{eqnarray}
and for $\hat{A}_{2\vk}$
\begin{eqnarray}
z \dla A_{2\vk} | B^\dagger \dra &=& \la A_{2\vk} | B^\dagger \ra 
+(\epsilon_\vk-\lambda_{2\vk}) \bar{n}_{\downarrow} \dla A_{1\vk} | B^\dagger \dra 
\nonumber \\
&+& 
(U + \lambda_{2\vk} )  \dla A_{2 \vk} | B^\dagger \dra + 
\dla A'_{3 \vk} | B^\dagger \dra
\end{eqnarray}
Consequently the new evolution given by the partial orthogonalization procedure is
\begin{equation}
K(\vk) =
\begin{pmatrix}
\epsilon_\vk & U \\ 
\bar{n}_{\downarrow}(\epsilon_\vk-\lambda_{2\vk})& U+\lambda_{2\vk}
\end{pmatrix}.
\end{equation}
The physical condition in Eq.~\eqref{phys-cond} is now satisfied for this evolution for any choice of the model parameters $\lambda_{2\vk} ,\bar{n}_{\downarrow},U,\epsilon_\vk$.
The approximate Green's function for the truncated theory becomes
\begin{equation}
G (z,\vk) = \frac{1}{z\eye -K(\vk)} N .
\end{equation}
Assuming no spin symmetry breaking, the parameter $\bar{n}_{\downarrow}$ can be determined self-consistently, by applying the fermionic characterization of the spectral theorem stated in  Eq.~\eqref{form:eq2} to  $G_{11}$, obtaining:
\begin{eqnarray}
\langle c^\dagger_{\vk\uparrow}c^{\,}_{\vk\uparrow} \rangle &=& \frac{1}{2\pi i} \oint dz f(z) G_{11}(z,\vk),
\\
\bar{n}_{\downarrow}= \bar{n}_{\uparrow} &=& \frac{1}{N_s} \sum_\vk \langle c^+_{\vk\uparrow}c^{\,}_{\vk\uparrow} \rangle .
\end{eqnarray}
To see that we can always write the residual Green's function highlighting the approximation involved in the truncation of the equation of motion let us analyze the special case where $\lambda_{2\vk}=\epsilon_\vk$.
The equation of motion of the Green's function with $B^\dagger  = c^\dagger_\vk$ becomes:
\begin{subequations}
\begin{eqnarray}
(z-\epsilon_\vk) \dla A_{1\vk} | c^{\dagger}_\vk \dra
&=&
1 +  U \dla A_{2\vk}| c^{\dagger}_\vk \dra,
\\
(z-\epsilon_k-U)\dla A_{2\vk}| c^{\dagger}_\vk \dra 
&=&
\bar{n}_{\downarrow} +  \dla A'_{3\vk} | c^{\dagger}_\vk \dra.
\end{eqnarray} 
\end{subequations}
Recalling that $A_{1\vk}=c_{\vk\uparrow}$ we find that the conventional fermion Green's function may be written exactly as
\begin{eqnarray}
\label{cc_cor}
\dla c^{\,}_{\vk\uparrow}| c^{\dagger}_{\vk\uparrow} \dra
&=& 
\frac{1-\bar{n}_{\downarrow}}{z-\epsilon_\vk} + \frac{\bar{n}_{\downarrow}}{z-\epsilon_\vk-U} 
\nonumber \\
& & +
\frac{U \dla A'_{3\vk} | c^{\dagger}_{\vk\uparrow} \dra}{(z-\epsilon_\vk-U)(z-\epsilon_\vk)}.
 \end{eqnarray}
From this it is clear that truncating the equation of motion implies that the term on the last line is neglected, making the approximation evident.
Moreover we note that  $\dla A'_{3\vk} | c^{\dagger}_{\vk\uparrow} \dra$ does not contain poles at $\epsilon_\vk$ and $\epsilon_\vk+U$ (since double poles in the original Green's function are not allowed) and its total spectral weight is vanishing (since $\la A'_{3\vk} | c^{\dagger}_{\vk\uparrow} \ra =0$). 
We may also note that excitations at $\epsilon_\vk$ and $U + \epsilon_\vk$ appears in the exact thermal Green's function (although their weight may be exponentially small) since they are exact energy differences between states with charge $1$ and $0$ and $2N_s -1$ and $2N_s$ respectively.

%
%
\section{The Global constraints on the  two-pole approximation of the Hubbard model}
\label{sec:sum_rules}
As noticed and stressed  by Mancini and Avella \cite{Mancini_Avella_1} the Roth procedure does not ensure that global sum rules such as those related to the Pauli principle and Ward identities are satisfied.
In the context of a two-pole approximation of the Hubbard model, these violations can be related to global constraint between averages. 
In particular the average double occupancy of the system can be evaluated in two inequivalent ways
\begin{eqnarray}
D =  \frac{1}{N_s} \sum_i \langle n_{i\downarrow} n_{i\uparrow} \rangle 
&=& \frac{1}{N_s} \sum_\vk \langle A_{1\vk}^\dagger A_{2\vk} \rangle 
\nonumber \\
&=& \frac{1}{N_s} \sum_\vk \langle A_{2\vk}^\dagger A_{2\vk} \rangle .
\end{eqnarray} 
At the operatorial level these two ways of writing the averages are equal.
Consequently when we evaluate these averages using the spectral theorem and the effective evolution, we have to make sure that
\begin{equation}
\label{const:global}
\Delta=\sum_\vk \frac{1}{2 \pi i } \oint dz 
\bigl[ G_{12}(z,\vk) - G_{22}(z,\vk) \bigr] f(z)=0 .
\end{equation}  
This constraint is very important, because it removes a fundamental ambiguity related to the determination of the energy in the Roth scheme.
In particular we can notice that in the previously studied solution, where we used $\lambda_{1\vk}=0$ and $\lambda_{2\vk}=\epsilon_\vk$ the constraint in Eq.~(\ref{const:global}) is automatically satisfied, because the argument of the integral is identically 0 for every $\vk$, making the solution suitable for unambiguous  physical interpretation. On a physical level this choice of the parameters makes the evolution diagonal in the two Hubbard operators which are orthogonal by construction.

%
%
\subsection{Variational determination of the orthogonalization parameter}
\label{sec:varpar}

As previously stated the determination of the orthogonalization parameters $\lambda_{2\vk}$ plays a crucial role.
Different values for this parameters gives different approximations to the true Green's function, all of them are physical in the sense that the spectral weights are positive and the excitation energies real, which is a fundamental requirement.
On the other hand different values of this parameter may correspond to quite different physics.
In some sense $\lambda_{2\vk}$ may be viewed as a kind of mean field parameter, in the sense of variational mean field theory.\cite{Chaikin_Lubensky_Book} Any choice for $\lambda_{2\vk}$ is allowed and gives physical results, but we want to determine the parameter to approximate the physics in the ``best'' possible way. The definition of ``best'' is however not unique, since approximations do not get everything correctly. Depending on what one choose to optimize different approximations will result.

It may be reasonable to demand that the solution posses the full lattice symmetry (i.e., assuming unbroken lattice symmetry). Then the evolution matrix $K(\vk)$ may be expanded in terms of proper basis functions with full lattice symmetry. The simplest non-trivial possibility is to take the ansatz for $\lambda_{2\vk}$ to be 
\begin{equation}
\lambda_{2\vk}= a_0 + a_1 \epsilon_\vk ,
\end{equation}
where $a_0$ and $a_1$ are some real $\vk$-independent constants. This may be viewed as the first two terms in a locality expansion.
 Let us also note that this is exactly the form for $\lambda_{2\vk}$ that is obtained in the Roth procedure in the two-pole approximation in the Hubbard model.\cite{Avella_2pole_1998}
 
If we further assume unbroken spin symmetry the average Free energy of the system (i.e. including the chemical potential term in the energy) may be evaluated using
\begin{equation}
\langle F \rangle = \sum_\vk \Big( 2(\epsilon_\vk-\mu)  \langle A_{1\vk}^ \dagger A^{\,}_{1\vk} \rangle 
+ U \langle A_{2\vk}^\dagger A^{\,}_{2\vk} \rangle \Big).
\end{equation}
To fix the parameters $a_0$ and $a_1$ we propose a zero temperature scheme based on minimizing the free energy. In particular we are going to use 
\begin{equation}
\Delta(a_0,a_1)=0, \qquad
\underset{a_0,a_1}{\text{min}} \langle F \rangle, 
\end{equation}
to fix $a_0$ and  $a_1$.
This is a constrained minimization problem and may be studied with standard methods in several ways. We can for example first fix $a_1$ and then try to solve the equation $\Delta(a_0,a_1) = 0$ for $a_0$. This may in general have more than one solution so it is crucial in this scheme to always check the number of roots of $\Delta(a_0)$. In addition the parameter $\bar{n}_\da$ will be determined self-consistently.
%
%
\section{Numerical results for the half filled case}
\label{half_fill_num}
In this section we are going to report some numerical results for the half filled case for a square lattice $100\times100$  at $T=0$. Throughout we will measure energies in units of $t$, which amounts to setting $t=1$. Half filling is obtained by taking $\mu = U/2$.
In Sec.~\ref{sec:analytical} below an analytical treatment of the half-filled case will be presented as well.

\begin{figure}[h!]
\includegraphics[width=0.4\textwidth]{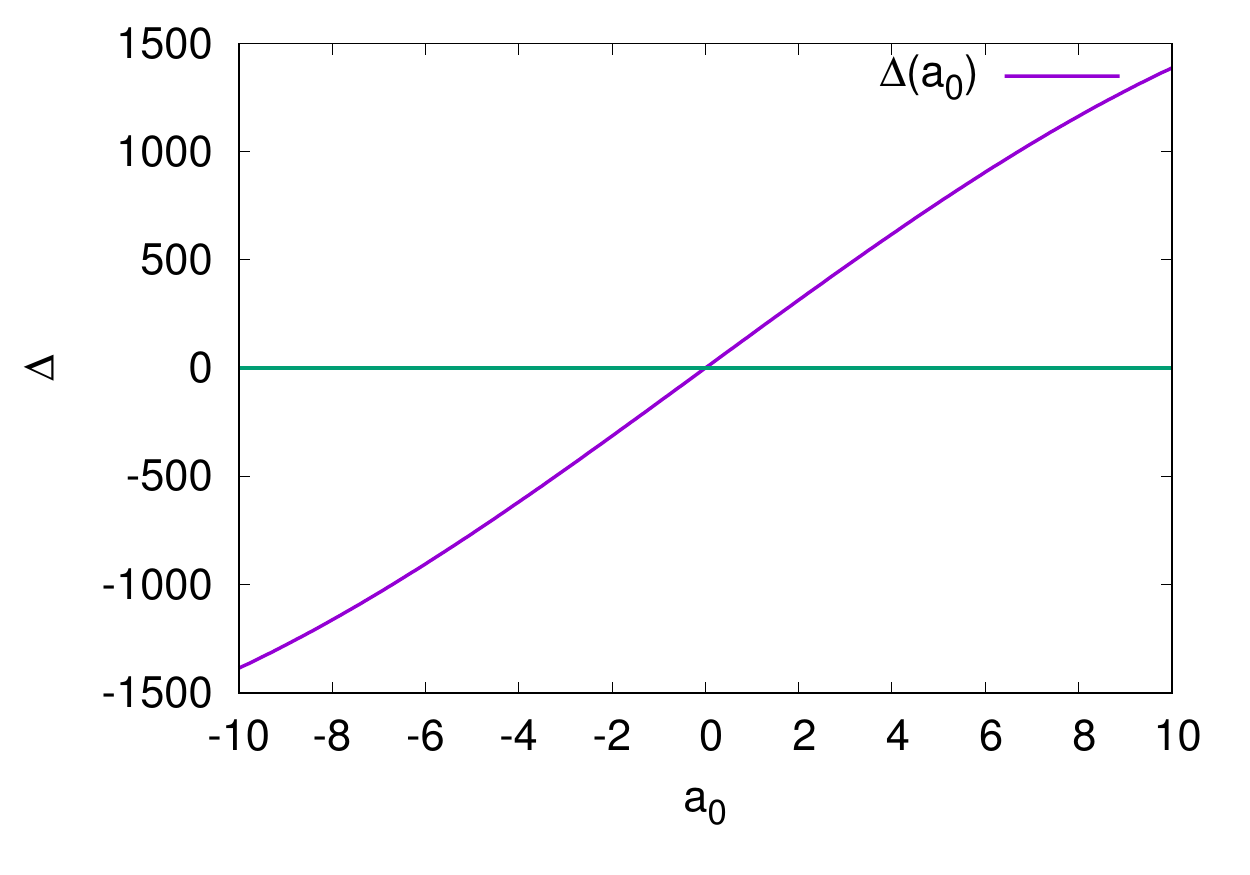} 
\caption{The function $\Delta(a_0)$ for parameters $U=12$, $a_1=-3$.}
\label{fig:delta_0} 
\end{figure}

In particular we are going to report the results obtained for two possible set of parameter $a_0$, $a_1$ which satisfy the constraint  Eq.~\eqref{const:global}: the $a_0=0$, $a_1=1$ case and 
the $a_0$, $a_1$ obtained by the variational scheme presented in Sec.~\ref{sec:varpar}. 
From Eq.~\eqref{const:global} it is possible to notice that for for $a_0=0$ we have $\Delta(0,a_1)=0$  independently on the value of $a_1$ and this is the only possible root, as can be seen in Fig.~\ref{fig:delta_0} (here we report $\Delta(a_0)$  only for a particular value of $a_1$ but the situation is the same for other values of $a_1$).

\begin{figure}
\includegraphics[width=0.4\textwidth]{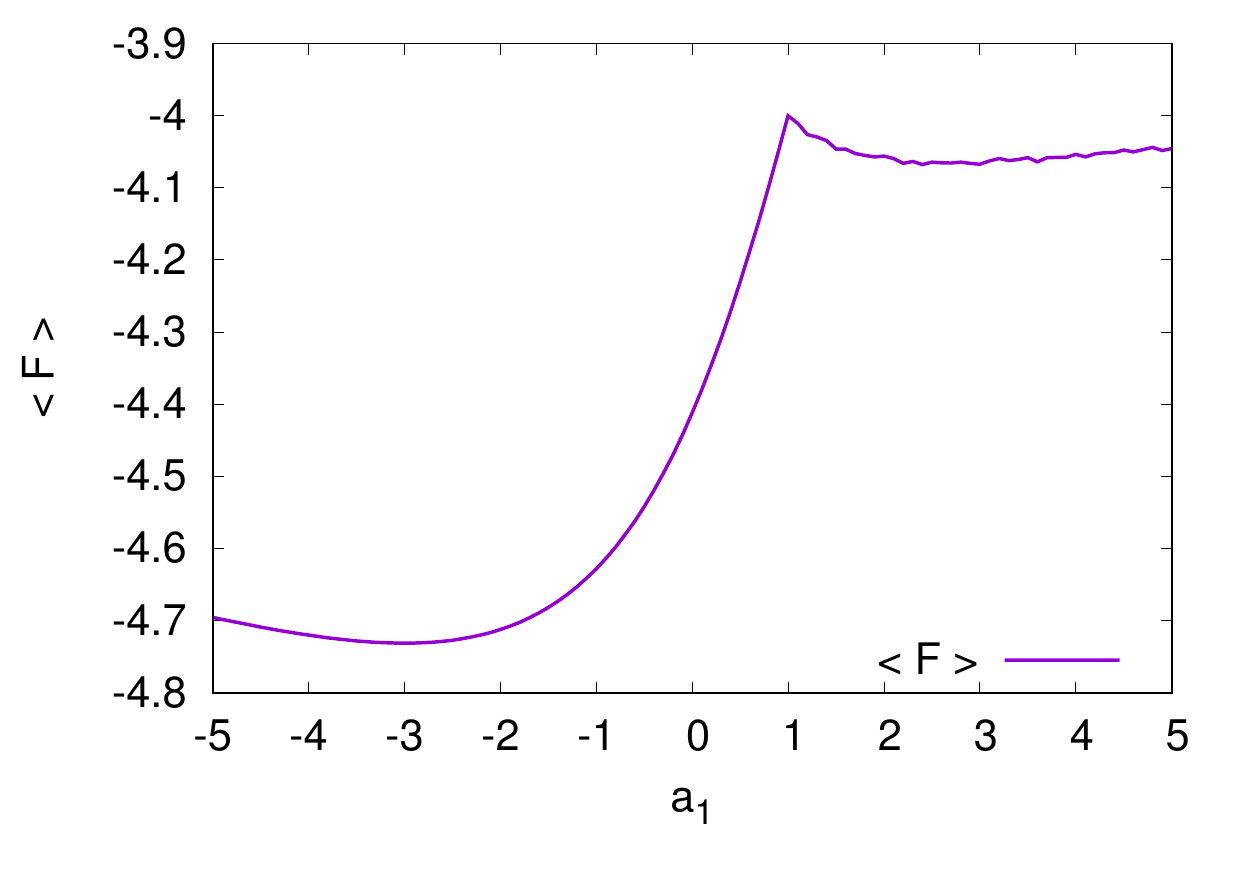} 
\caption{Expectation value of the Free energy $\la F \ra (a_1)$ for $U=8$, $a_0=0$. The global minimum near $a_1 = -3$ is clearly visible.}
\label{fig:fe_a1} 
\end{figure}

To carry out the Free energy minimization carefully, it is important to have a sketch of the Free energy landscape as a function $a_1$, since we will put $a_0=0$. A representative curve can be seen in Fig.~\ref{fig:fe_a1}, and we notice that $\langle F \rangle (a_1)$ posses a global minima for negative values of $a_1$. In particular after carry out the constrained minimization numerically we found that the minimum of the free energy is reached for $a_1=-3$, $a_0=0$, independently on the coupling strength $U$.
At this point we are going to compare the avarage energy and double occupancy obtained for the two choices of the decoupling parameter $a_1=1$, $a_0 =0$ and $a_1=-3$, $a_0=0$ against the benchmark results gathered from Le Blanc {\it et al.},\cite{LeBlanc2015} reported respectively in Tab.~\ref{table:zeroTE} and Tab~\ref{table:zeroTD}. From Tab.~\ref{table:zeroTE}   it is possible to notice both decouplings $a_1=1$ and $a_1=-3$ predicts energies that may be lower than the exact methods. Consequently this scheme is {\it not} variational in the usual sense. This is expected since we are approximating the Green's function. In fact, despite that we know analytically the term neglected in the Green's function Eq.~(\ref{cc_cor}), the approximate Green's function properties are determined self-consistently within the truncated theory which can be different from the original one.

From Tab.~\ref{table:zeroTD} we can notice that  in the case $a_1=1$ the double occupancy drop to zero for $U\ge8$. In the other case $a_1=-3$ double occupancy is predicted to be of the order $(t/U)^2$, which agree at least in order of magnitude with the benchmark results.

\begin{figure}[h!]
\includegraphics[width=0.5\textwidth]{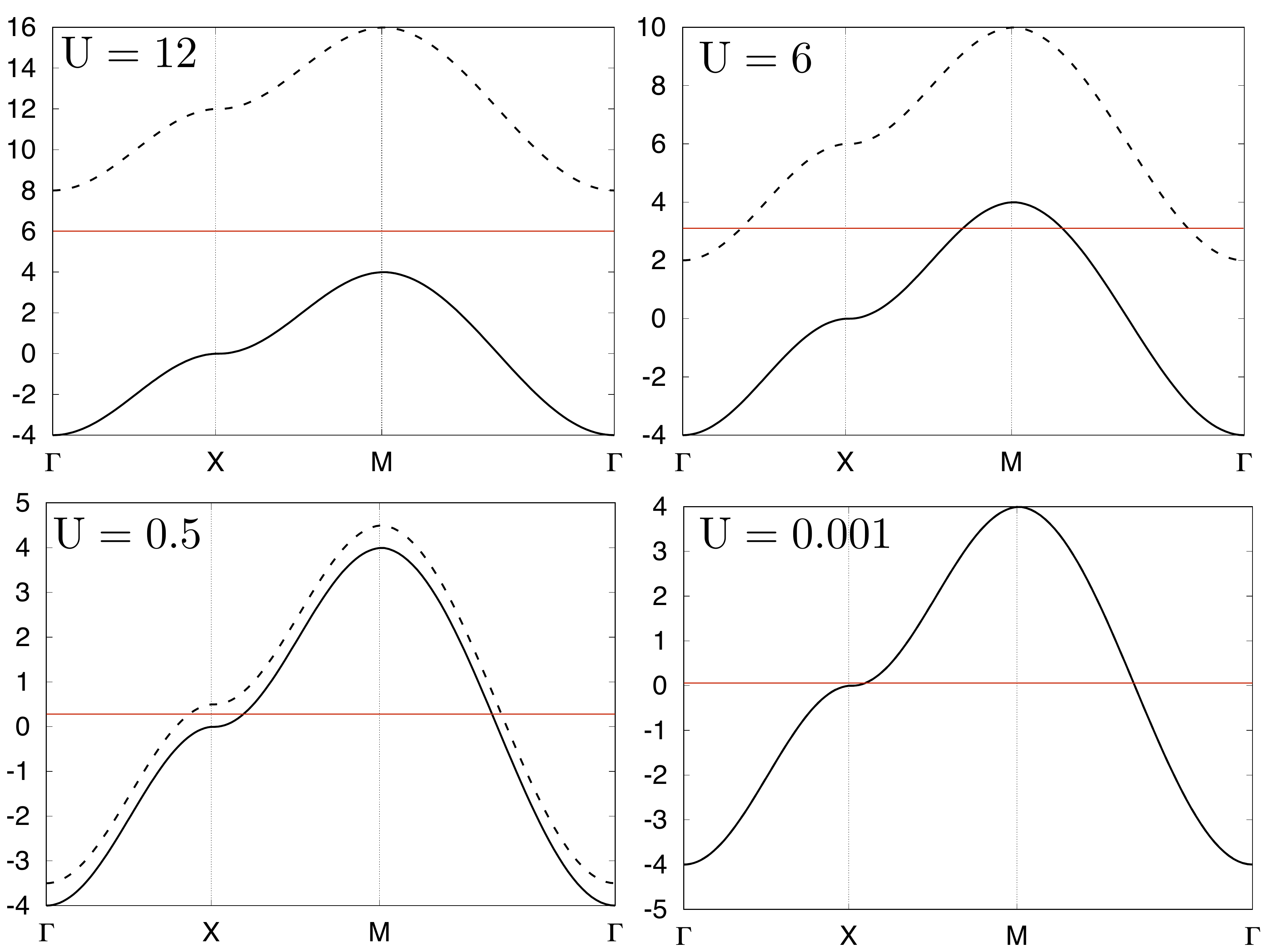} 
\caption{Band structures for different values of $U$ and $a_1=1$ at half-filling. The red line indicates the chemical potential.}
\label{fig:band_a1} 
\end{figure}

\begin{figure}[h!]
\includegraphics[width=0.5\textwidth]{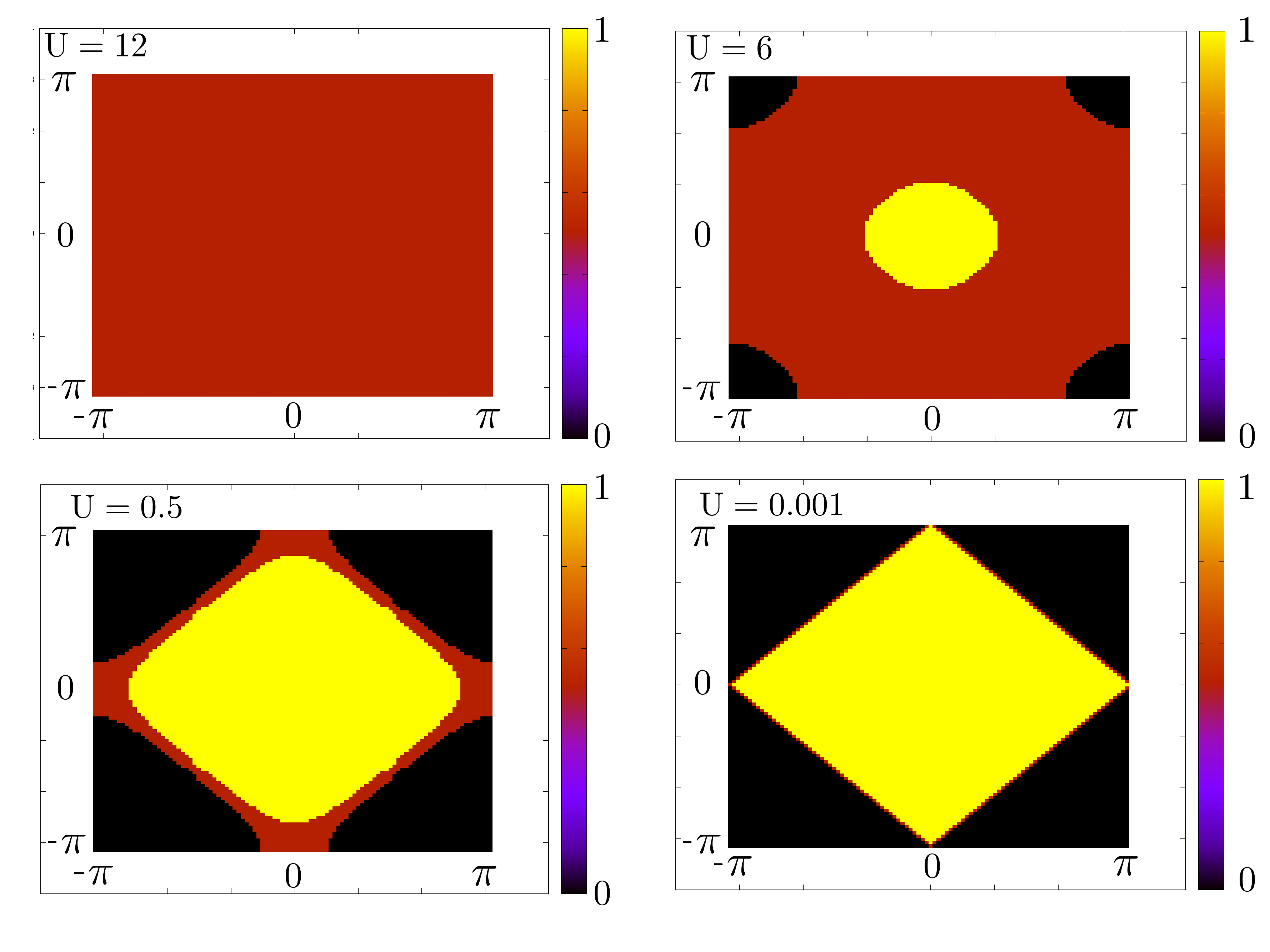} 
\caption{Colormap of the average occupation in the first Brillouin zone for different values of $U$ and $a_1=1$ at half-filling.}
\label{fig:brill_a1} 
\end{figure}

It is important to highlight that despite the crudeness of the two-pole approximation, this scheme independently on the value of parameter $a_1$ is capable to capture the effect of the correlation predicting a double occupancy that is significantly reduced from the mean field value $n_{\ua}n_\da =1/4$. 
The two possible choice of parameters $a_1=$, $a_0=0$ and $a_1=-3$, $a_0=0$ predict big differences at the level of predicted observables however. 
In particular for the case $a_0=0$, $a_1=1$ the truncated theory posses two energy bands shifted rigidly by $U$ (i.e., independently of the momentum), as may be seen in  Eq.~(\ref{cc_cor}) and in Fig.~\ref{fig:band_a1}.
With this choice of parameters the occupations of all the k-points in the first Brillouin zone are half occupied for $U>8t$ and for $U<8t$ we have a formation of fully occupied region around the $\Gamma$ point surrounded by a region of half occupied k-points, and an empty region close to $M$ point. The half-filled region in between shrinks as the interaction strength is decreased as can be seen in Fig.~\ref{fig:brill_a1}. We can also notice that in the limit $U \to 0$, we recover the diamond-shaped Fermi surface for free fermions on the square lattice at half-filling.

\begin{figure}[h!]
\includegraphics[width=0.5\textwidth]{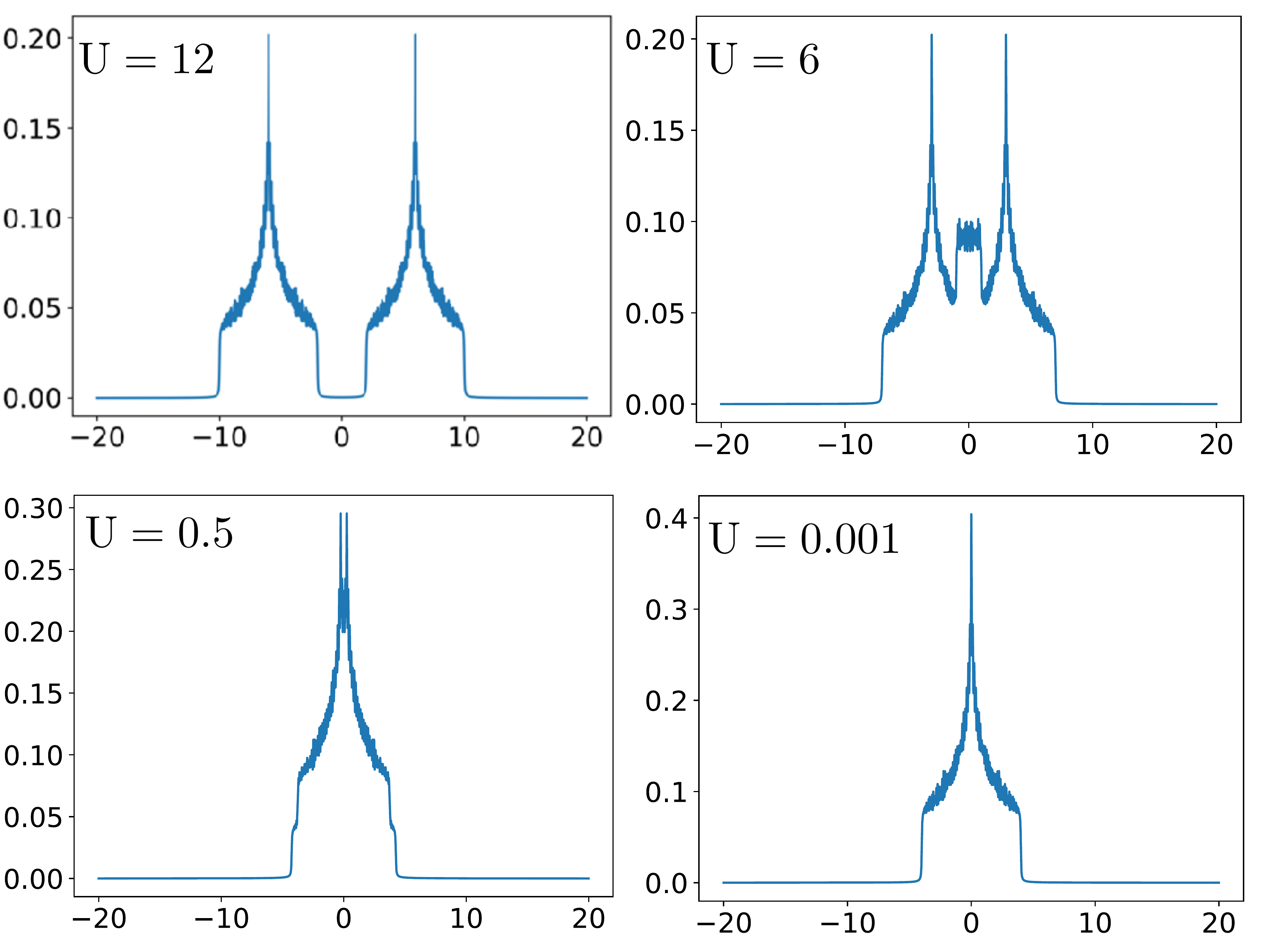} 
\caption{DOS for different values of $U$ and $a_1=1$ at half-filling. Energies on the $x$-axis are measured with respect to the chemical potential.}
\label{fig:dos_a1} 
\end{figure}

To capture the metallic or insulting behavior of the solution one should in principle evaluate the conductivity or the charge-charge correlation function, which is in principle unaccessible with the operators used here. 
However we can have an indication on the metallic or insulating behavior of the system by analyzing density of states. 
Let us first consider the case $a_1 = 1$. In this case we can see in Fig.~\ref{fig:dos_a1} that for $U>8t$ the density of states posses an hard gap and there is a formation of separated lower and upper Hubbard bands, which is a signature of an insulating phase.
For $U<8t$ the lower and upper Hubbard bands overlap giving rise to a gapless density of state which is an indication of a metallic phase.
Consequently for the choice of parameter $a_1=1$, $a_0=0$, we can notice that  $U=8$ represent a critical value of the interaction above which the system is in an insulating state and below which the system is in a metallic state.
 
\begin{figure}[h!]
\includegraphics[width=0.5\textwidth]{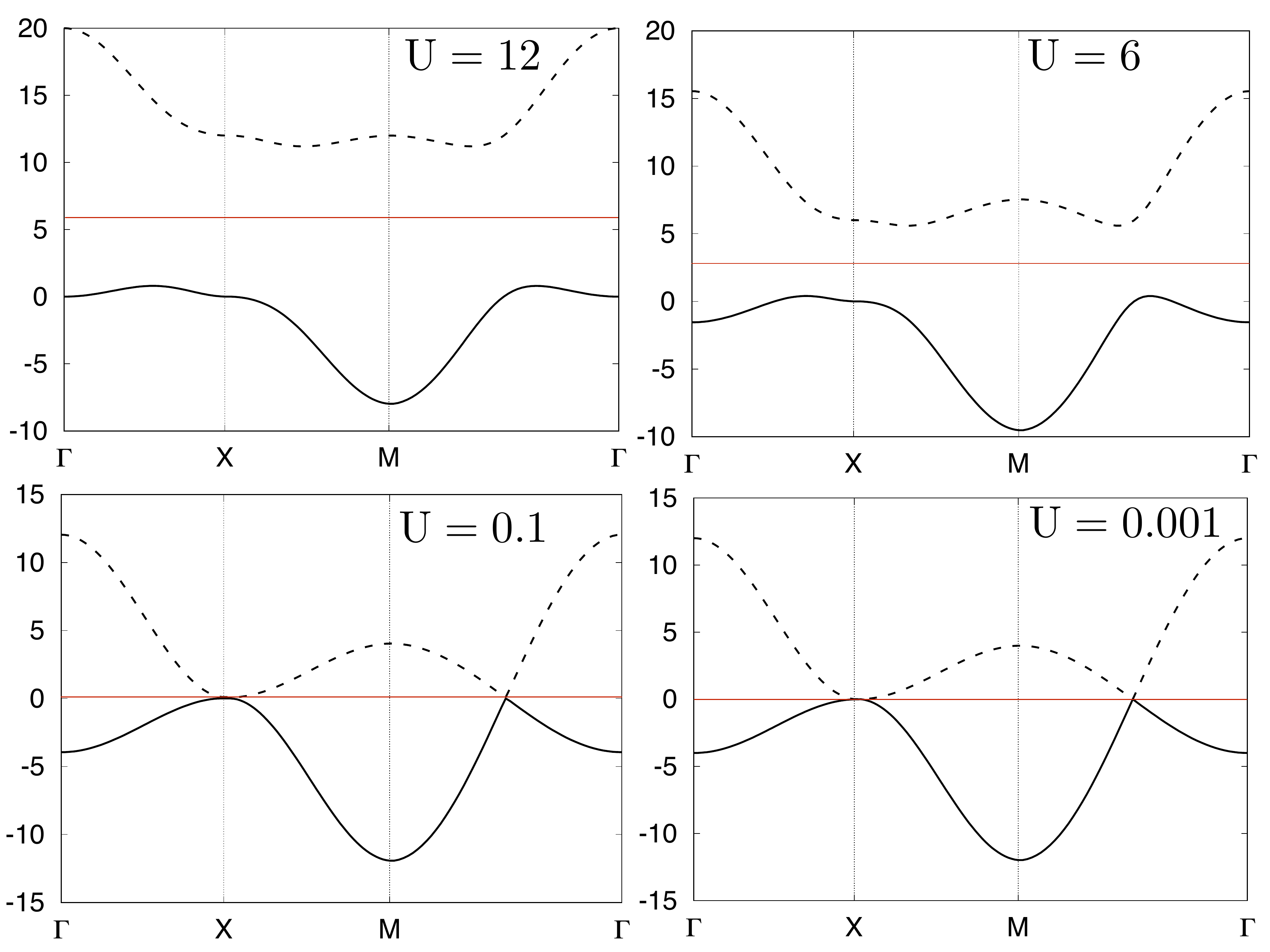} 
\caption{Band structures for different values of $U$ and $a_1=-3$ at half-filling. The red line indicates the chemical potential.}
\label{fig:band_a-3} 
\end{figure}

\begin{figure}
\includegraphics[width=0.5\textwidth]{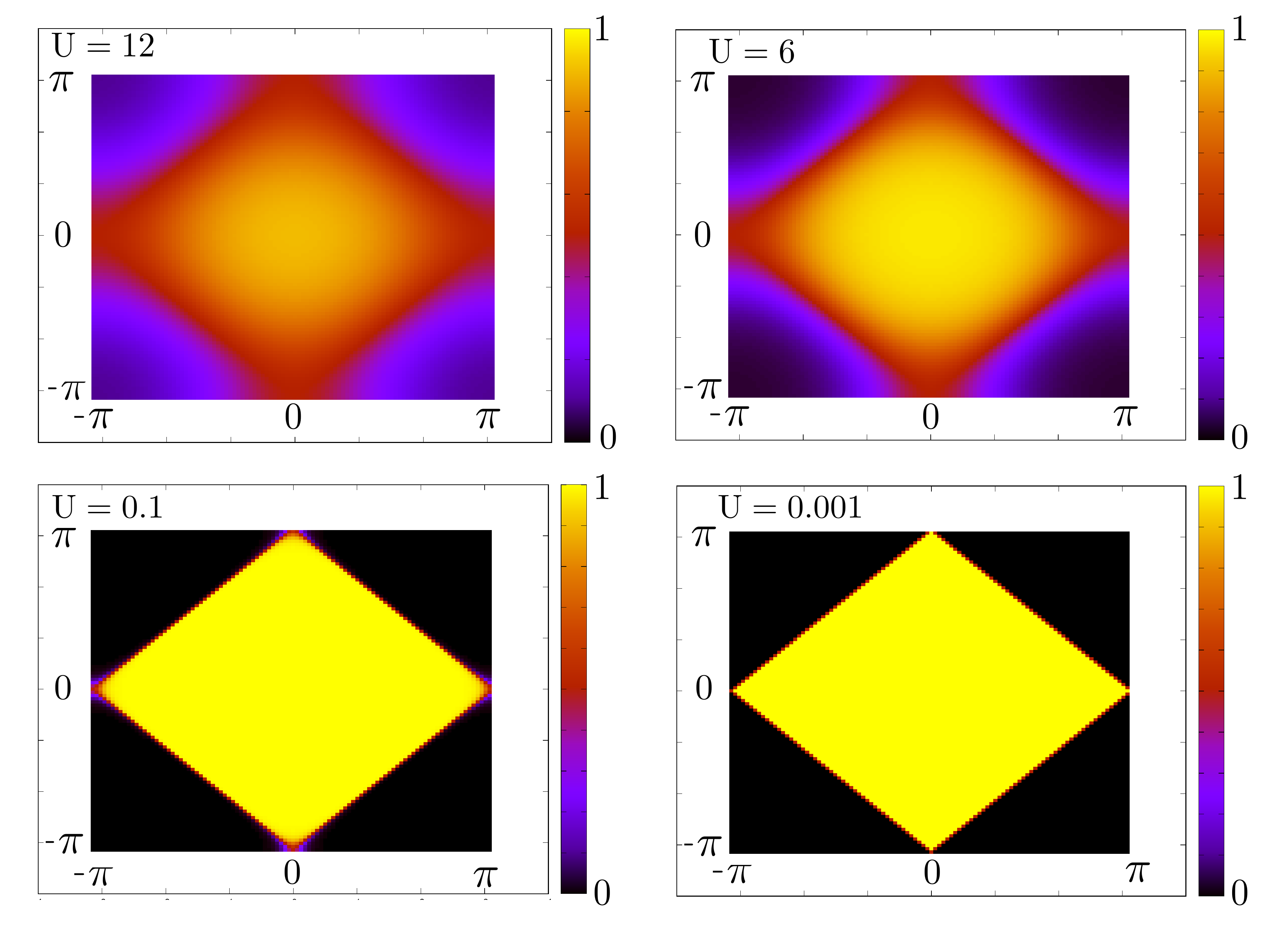} 
\caption{Colormap of the average occupation in the first Brillouin zone for different values of $U$ and $a_1=-3$ at half-filling.}
\label{fig:brill_a-3} 
\end{figure}

\begin{figure}[h!]
\includegraphics[width=0.5\textwidth]{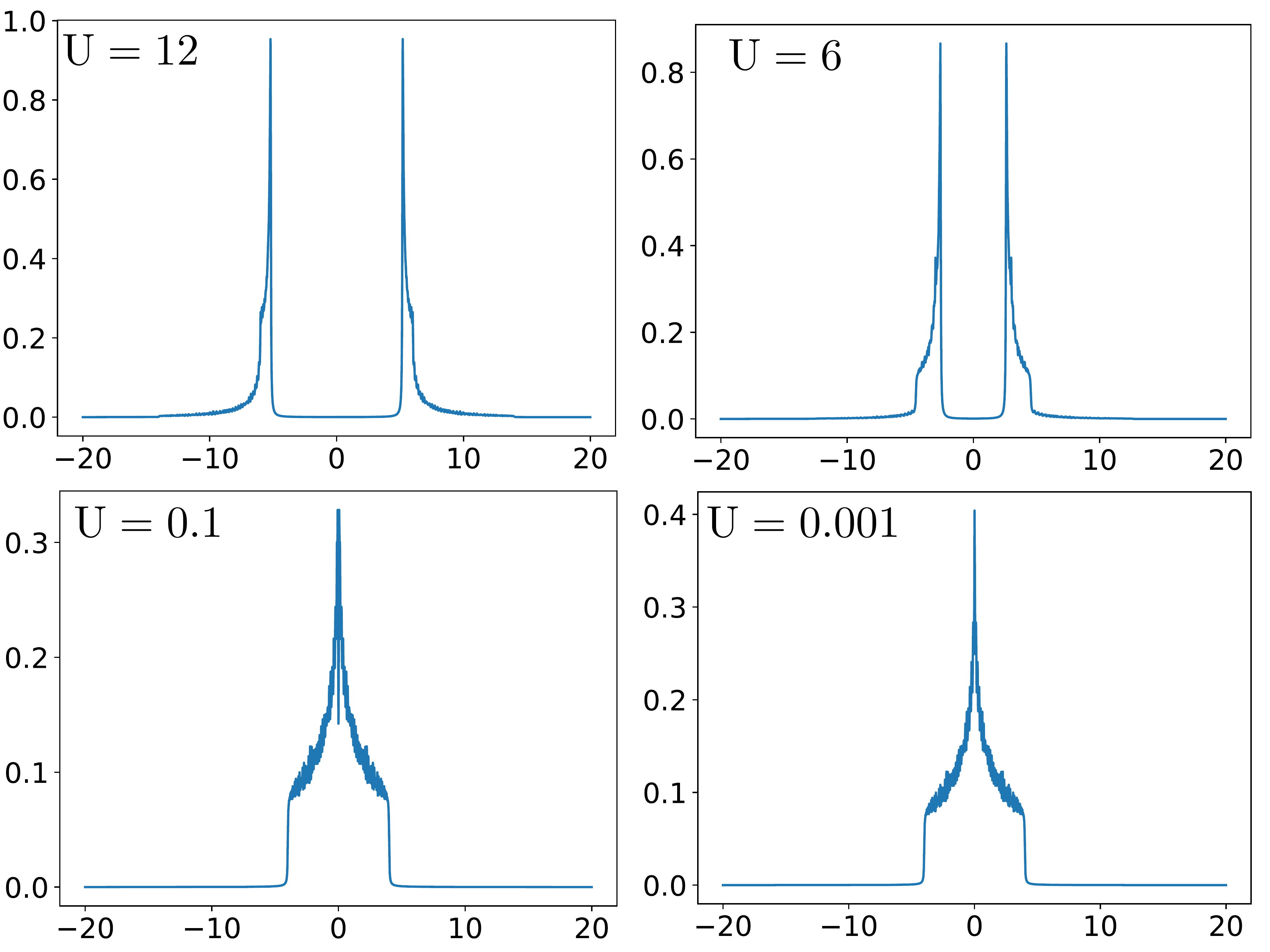} 
\caption{DOS for different values of $U$ and $a_1=-3$ at half-filing. Energies on the $x$-axis are measured with respect to the chemical potential. There is always a tiny gap that is not visible on this scale in the lower two panels.}
\label{fig:dos_a-3} 
\end{figure}

A radically different behavior is predicted by the choice of parameters $a_1=-3$, $a_0=0$. In this case the system posseses two bands that repel with increasing interaction strength and there is always a small gap between the two bands for any non-vanishing value of $U$. The gap becomes very small for small interactions as can be seen by looking at the case $U=0.1$ in Fig.~\ref{fig:band_a-3}.
With the choice of these parameters the occupation in first Brillouin zone is characterized by the presence of an almost fully occupied region around the $\Gamma$ point which changes continuously to a low but non-zero occupation at the corner of the first Brillouin zone (the M point). As the interaction is decreased the almost fully occupied region around the $\Gamma$ becomes increasingly occupied and the corner of the Brillouin zone get increasingly depleted. From the figures it looks like a Fermi surface is formed at a $U=0.1$ along the high symmetry vector $M\Gamma$ as it is possible to notice in Fig.~\ref{fig:brill_a-3}. There is however a small gap that is not seen on this scale, this becomes clear in the analytic treatment in Sec.~\ref{sec:analytical}. In the limit of $U\to0$ also in this case we recover the diamond-shaped Fermi surface for a free electron gas on square lattice.
As we did for the case $a_1=1$, $a_0=0$ we can also study the density of state in order to get an indication on the phase of the system. In this case there is always a gap between the upper and lower Hubbard band, but the gap is very small for small $U$ as can be seen in Fig.~\ref{fig:dos_a-3}.
However in order to better characterize the possible phases of the system for this choice of parameters it is going to be beneficial a study of the $\langle F \rangle(U)$. In fact, the solution with $a_1 = -3$ always has a lower expectation value of the Free energy than the solution at $a_1 = 1$. Moreover, since $a_1 = -3$ is insulating, our scheme indicates that the insulator is stable at half-filling.

\subsection{Relation to the two-pole approximation of Avella and collaborators}

We can relate our approach to that of Avella et al by comparing the associated $E$-matrices.\cite{Avella_2pole_1998}
The relation between their parameters ($\Delta$ and $p$) and ours ($a_0$ and $a_1$) are given by
\begin{subequations}
\begin{eqnarray}
-2dt \Delta  &=& \bar{n}_\da (1- \bar{n}_\da) a_0   ,
\\
p &=& \bar{n}_\da (1- \bar{n}_\da)  a_1 + \bar{n}^2_\da .
\end{eqnarray}
\end{subequations}
The issue of the determination of these parameters is discussed at length in Ref.~\onlinecite{Avella_2pole_1998}.
We note that they choose to determine $\Delta$ self-consistently from the Green's function, and fix $p$ so that Pauli principle is satisfied. This is different from our procedure where $a_0$ (and therefore $\Delta$) is determined so that the Pauli principle is satisfied, in the next step we fix $a_1$ to minimize expectation value of the Free energy.
Comparing the results we also have two classes of solutions, but the parameters obtained are not identical.

\section{Analytical results}
\label{sec:analytical}

Since the two-pole approximation involves $2 \times 2$ matrices everything may be evaluated exactly.
A straightforward calculation gives (dropping $\vk$ indexes on $\epsilon_\vk$ and $\lambda_{2\vk}$ and other parameters for brevity)
\begin{subequations}
\begin{eqnarray}
\dla c^{\,}_{\vk \uparrow} | c^{\dagger}_{\vk \uparrow} \dra 
&=& 
\frac{1}{2} \Bigl( \frac{1+\delta_1}{z-E_-} + \frac{1-\delta_1}{z-E_+} \Bigr) ,
\\
\dla \eta^{\,}_{\vk \uparrow} | \eta^{\dagger}_{\vk \uparrow} - c^{\dagger}_{\vk \uparrow}  \dra
&=& 
\delta_2 \Bigl( \frac{1}{z-E_-} - \frac{1}{z-E_+} \Bigr),
\\
\dla \eta^{\,}_{\vk \uparrow} | \eta^{\dagger}_{\vk \uparrow} \dra
&=&
\frac{\bar{n}_\da}{2} \Bigl(
\frac{1+\delta_3}{z-E_-} + \frac{1-\delta_3}{z-E_+}
\Bigr),
\end{eqnarray}
\end{subequations}
where the poles are located at
\begin{equation}
E_{\pm} = \frac{U+\epsilon + \lambda_2 \pm 
\sqrt{(U - \epsilon + \lambda_2)^2 + 4 \bar{n}_\da U (\epsilon-\lambda_{2})}}{2}.
\end{equation}
The other parameters that are related to the weight of the poles are
\begin{subequations}
\begin{eqnarray}
\delta_1 &=& \frac{U(1-2 \bar{n}_\da) + \lambda_2 - \epsilon}{\sqrt{(U - \epsilon + \lambda_2)^2 + 4 \bar{n}_\da U (\epsilon-\lambda_{2})}},
\\
\delta_2 &=& \frac{\bar{n}_\da (1-\bar{n}_\da) (\epsilon - \lambda_2)}{\sqrt{(U - \epsilon + \lambda_2)^2 + 4 \bar{n}_\da U (\epsilon-\lambda_{2})}},
\\
\delta_3 &=& - \frac{U + (1 - 2 \bar{n}_\da) (\lambda_2 - \epsilon)}{\sqrt{(U - \epsilon + \lambda_2)^2 + 4 \bar{n}_\da U (\epsilon-\lambda_{2})}} .
\end{eqnarray}
\end{subequations}
Using this we may calculate many quantities of interest, such as the density of spin-up electrons
\begin{equation}
\bar{n}_\ua = \frac{1}{N_s} \sum_\vk 
\Bigl(
\frac{1+\delta_{1\vk}}{2}n_{-\vk} + \frac{1-\delta_{1\vk}}{2} n_{+\vk} \Bigr),
\label{eq:analytic_density}
\end{equation}
the Pauli principle constraint ($\Delta = 0$)
\begin{equation}
\sum_\vk \delta_{2\vk} (n_{- \vk}  - n_{+ \vk} ) = 0 ,
\end{equation}
and average double occupancy
\begin{equation}
D =  \bar{n}_\da \frac{1}{N_s} \sum_\vk 
\Bigl( \frac{1+\delta_{3\vk}}{2}n_{- \vk} + \frac{1-\delta_{3\vk}}{2} n_{+ \vk} \Bigr) ,
\end{equation}
as well as the average kinetic energy (of two spin species)
\begin{equation}
\la \hat{H}_0 \ra = \frac{2}{N_s} \sum_\vk \epsilon_\vk
\Bigl(
\frac{1+\delta_{1\vk}}{2}n_{- \vk} + \frac{1-\delta_{1\vk}}{2} n_{+ \vk} \Bigr).
\end{equation}

\subsection{Simplifying assumptions -- insulator}

It is possible to find the solution with $a_1 = -3$ obtained in the numerical study above analytically. In this subsection we present this solution is some detail since it provides an interesting zeroth order approximate Green's function at half-filling.

Let us assume that $U$ is sufficiently large and chemical potential sufficiently small so that $n_{-\vk} = 1$ and $n_{+\vk} = 0$ for all $\vk$. We must then have
\begin{equation}
\frac{1}{N_s} \sum_\vk \delta_{2\vk} = 0 .
\end{equation}
Then $\bar{n}_\ua = \bar{n}_\da = 1/2$ solves Eq.~\eqref{eq:analytic_density}. With this choice
\begin{eqnarray}
\delta_2 = \frac{1}{4} \frac{\epsilon - \lambda_2}{\sqrt{U^2 + (\epsilon - \lambda_2)^2}},
\end{eqnarray}
and therefore any $\lambda_2 = a_1 \epsilon$ will satisfy the Pauli principle constraint.
The other parameters then become
\begin{subequations}
\begin{eqnarray}
\delta_1 &=& \frac{ (a_1 -1 ) \epsilon}{\sqrt{U^2 + (a_1 - 1)^2 \epsilon^2}},
\\
\delta_3 &=& - \frac{U }{\sqrt{U^2 + (a_1 - 1)^2 \epsilon^2}}.
\end{eqnarray}
\end{subequations}
Using this me may write down expressions for average double occupancy
\begin{equation}
D =  \frac{1}{4} \frac{1}{N_s} \sum_\vk 
\Bigl(1- \frac{U }{\sqrt{U^2 + (a_1 - 1)^2 \epsilon_\vk^2}} \Bigr) ,
\end{equation}
and average kinetic energy
\begin{equation}
\la \hat{H}_0 \ra = \frac{1}{N_s} \sum_\vk 
 \frac{ (a_1 -1 ) \epsilon^2_\vk}{\sqrt{U^2 + (a_1 - 1)^2 \epsilon_\vk^2}} .
\end{equation}
Minimizing $\la \hat{H}_0 \ra + U D$ we find a minimum at $a_1 = -3$, with the energy being
\begin{equation}
\la \hat{H}_0 \ra + U D = \frac{1}{4} \frac{1}{N_s} \sum_\vk 
\Bigl( U - \sqrt{U^2 + (4 \epsilon_\vk)^2} \Bigr).
\end{equation}
The gain in energy due to hopping is increased with respect to more conventional approaches, such as anti-ferromagnetic mean field.
Let us also note that the band structure for this solution is
\begin{equation}
E_{\pm} = \frac{U  -2 \epsilon_\vk \pm \sqrt{U^2 + (4 \epsilon_\vk)^2}}{2} ,
\end{equation}
in the large-$U$ limit we therefore get
\begin{equation}
E_{\pm} \approx  U \Bigl( \frac{1\pm 1}{2}\Bigr) - \epsilon_\vk,
\end{equation}
giving us two Hubbard bands with the full bare non-interacting bandwidth. Note however that the sign of the kinetic term is opposite to what if would be in the non-interacting case.
The solution $a_0 = 0$, $a_1 = 1$ in the same region has $D=0$ and $\la \hat{H}_0 \ra=0$ so is always higher in energy than $a_0 = 0, a_1 = -3$. This agrees with our numerical findings.

\section{Hole doped case in the strong coupling regime} 
\label{sec:holedoped_12}

\begin{figure}[h!]
\includegraphics[width=0.5\textwidth]{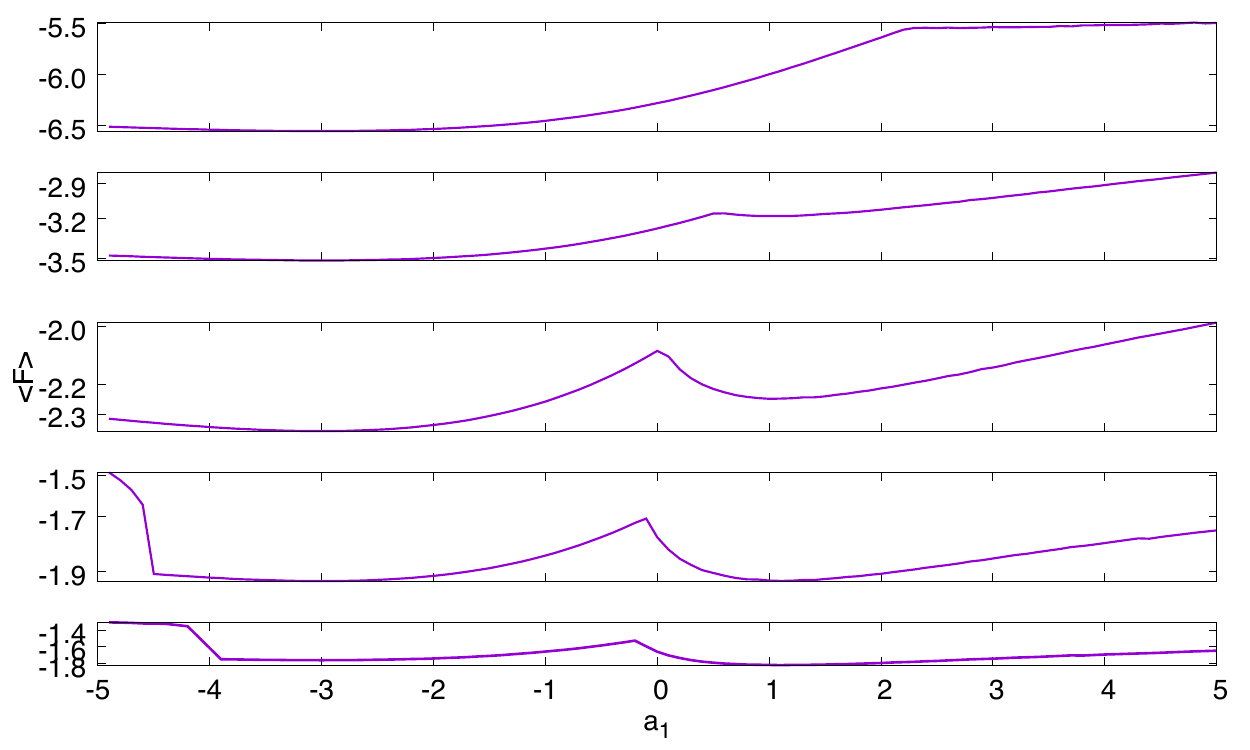} 
\caption{Free energy expectation value $\langle F \rangle $ as a function of $a_1$ for different values of the chemical potential decreasing from the top panel $\mu \in \{6.0,3.0,1.8,1.4,1.2\}$, interaction strength is $U=12$.}
\label{fig:min_shiftU12} 
\end{figure}

\begin{figure}[h!]
\includegraphics[width=0.5\textwidth]{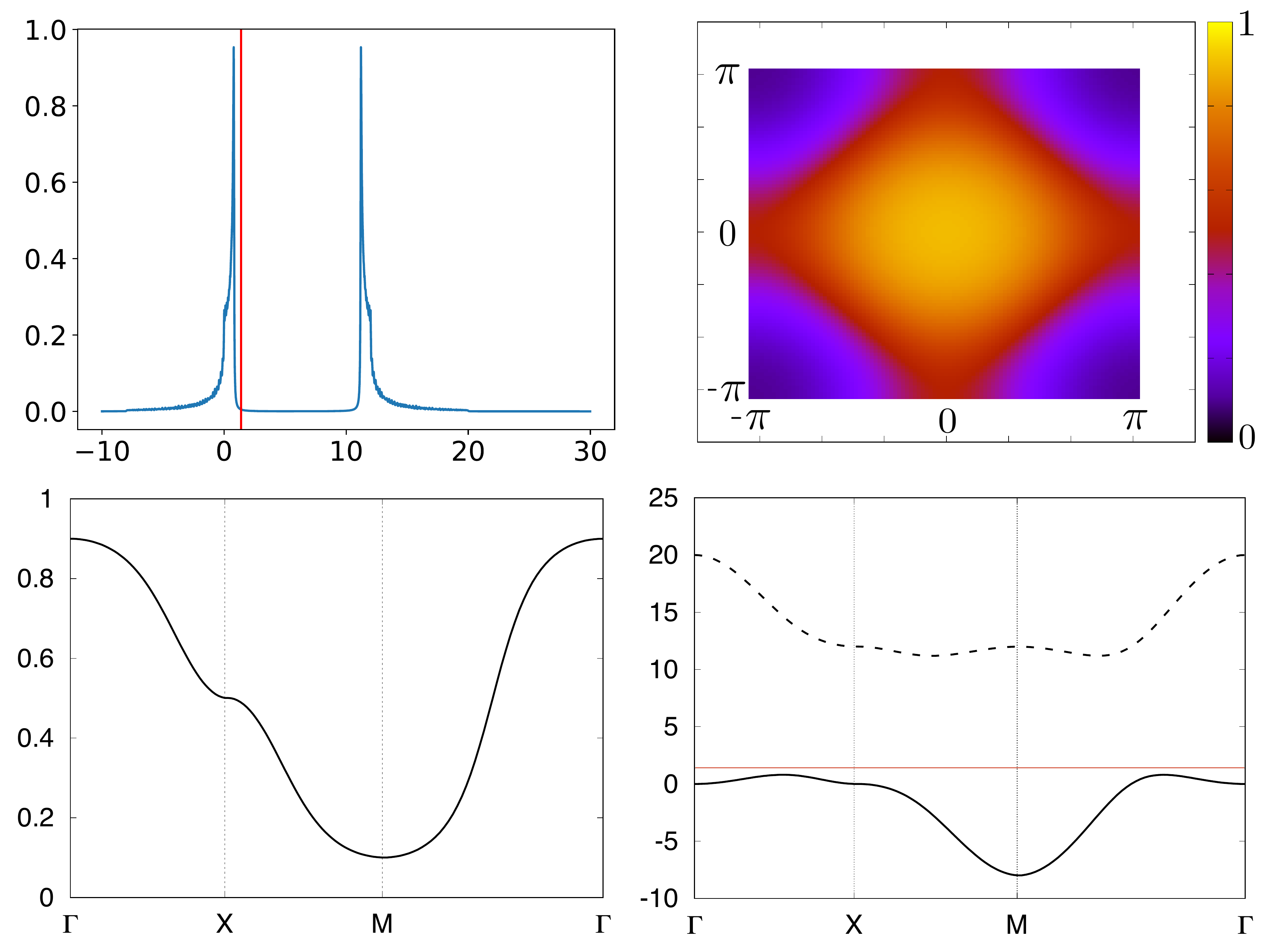} 
\caption{DOS (top left), colormap of the average occupation in the first Brillouin zone (top right),
average occupations along high symmetry lines (bottom left), 
and band structure (bottom right), for $U=12$ , $\mu=1.4$ and $a_1=-3$ and  $\langle n_{\downarrow} \rangle=0.5$ .
The red line indicates the chemical potential.}
\label{fig:DOS_hardgap} 
\end{figure}

\begin{figure}
\includegraphics[width=0.5\textwidth]{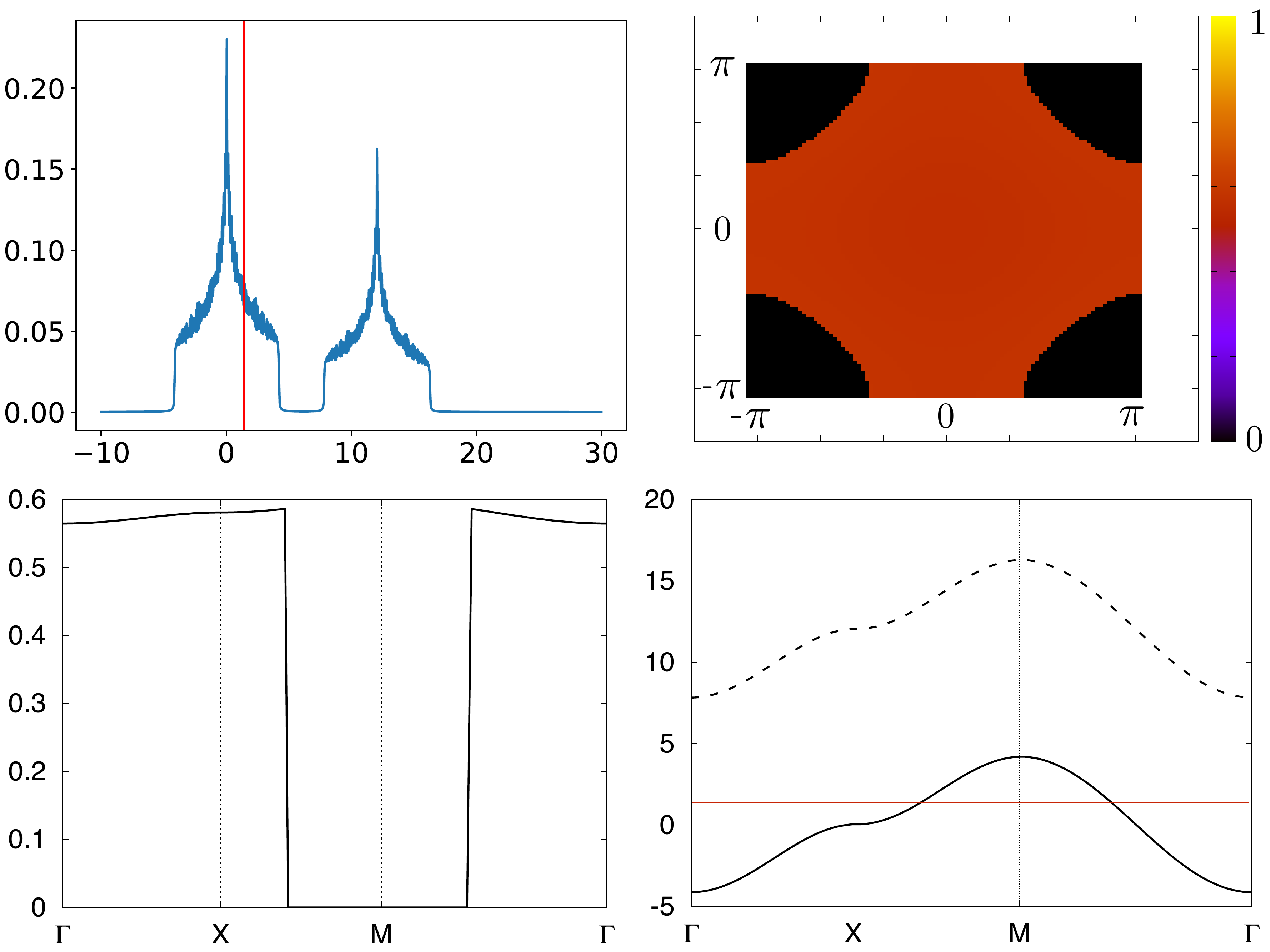} 
\caption{DOS (top left), colormap of the average occupation in the first Brillouin zone (top right),
average occupations along high symmetry lines (bottom left), 
and band structure (bottom right), for $U=12$ , $\mu=1.4$ and $a_1=1.1$ and  $\langle n_{\downarrow} \rangle=0.42$.
The red line indicates the chemical potential.}
\label{fig:DOS_mildgap} 
\end{figure}

In this section we are going to apply our scheme in the hole doped case for an interaction strength  larger than the bandwidth namely $U=12$. From the Free energy plots in Fig.~\ref{fig:min_shiftU12} it is clear that upon hole doping (decreasing the chemical potential) the minima around $a_1=1$ is pushed down in energy with respect to the one near $a_1 = -3$, until it becomes the global one below a critical value near $\mu = 1.4$.
From an analysis of the DOS in Fig.~\ref{fig:DOS_hardgap}, it is possible to notice that the solution around $a_1=-3$  is characterized by the presence of an hard gap and the system is predicted to be insulating  up to $\mu=1.4$.
On the other hand the solution around $a_1=1$ is characterized by a smaller gap and  when $\mu<1.4 $, becomes the global minima of the Free energy. 
The DOS found in Fig.~\ref{fig:DOS_mildgap} suggests the formation of a metallic phase  and it is possible to notice a spectral weight transfer from high energy states to the low energy ones. 

\begin{figure}
\includegraphics[width=0.5\textwidth]{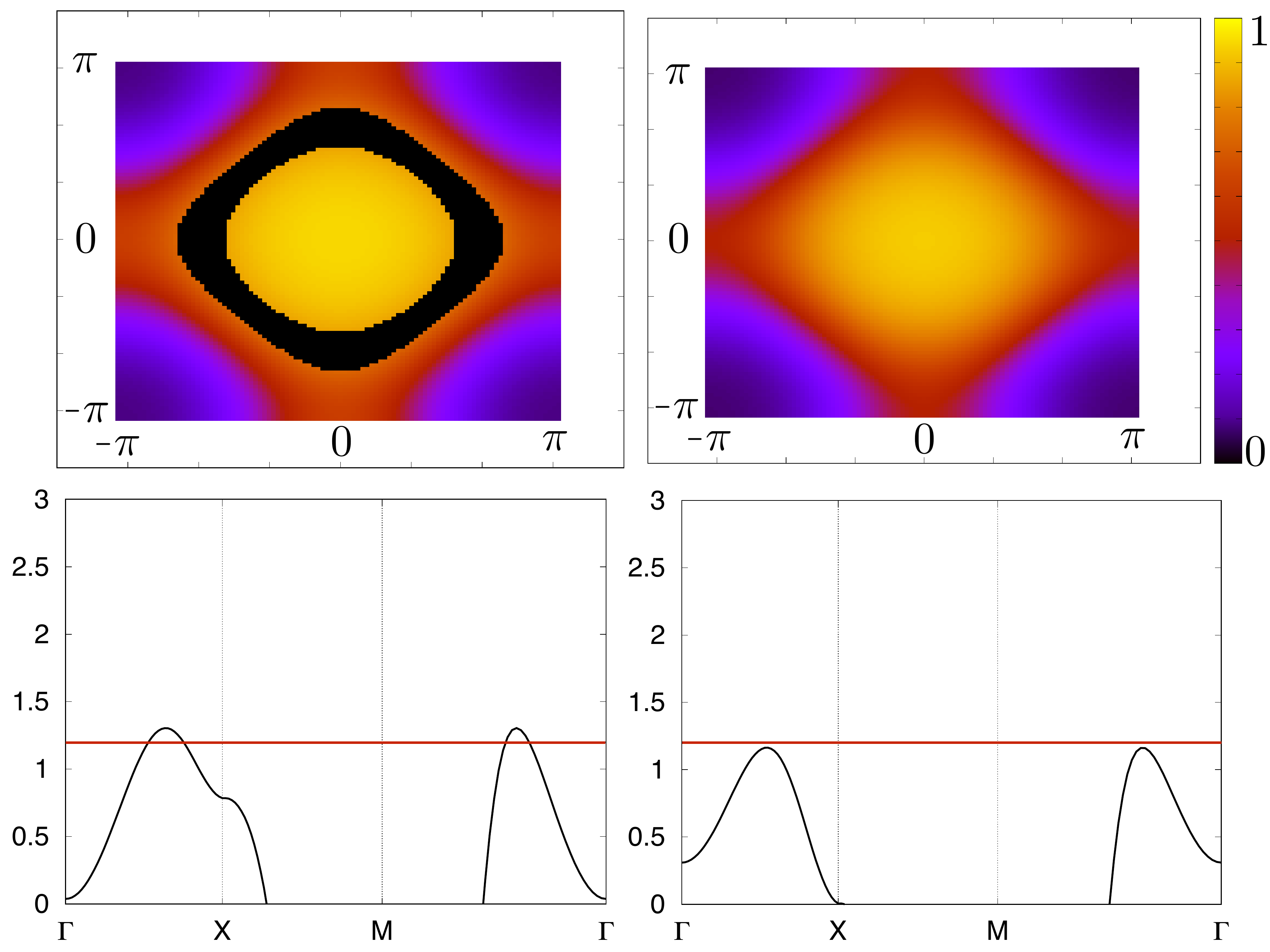} 
\caption{Colormap of the occupation in the first Brillouin zone for $U=12$ and $\mu=1.2$, for $a_1=-4.1$ (top left) and $a_1=-3.9$ (top right). Zoom in of band structure for the corresponding two cases (lower two panels) making the Lifshitz transition apparent. The red line indicates the chemical potential.}
\label{fig:comparison_lifshitz} 
\end{figure}

Another interesting feature of Fig.~\ref{fig:min_shiftU12} for $\mu \le 1.4$ is that the Free energy as a function of $a_1$ features a discontinuous behavior at some values in the range $a_1\in[-5,-4]$.
In this case the lower band get attracted to the upper one, pushing it above the chemical potential for certain momenta. This results in the formation of unoccupied k-points in the first Brillouin and two Fermi surfaces that may be seen in Fig.~\ref{fig:comparison_lifshitz}. This may be viewed as a Lifshitz transition.\cite{Lifshitz1960,Slizovskiy2018}
This solution is however not energetically favorable, and is not likely stabilized without additional interactions.

\section{Hole doped case in the intermediate coupling regime} 
\label{sec:holedoped_4}

\begin{figure}
\includegraphics[width=0.5\textwidth]{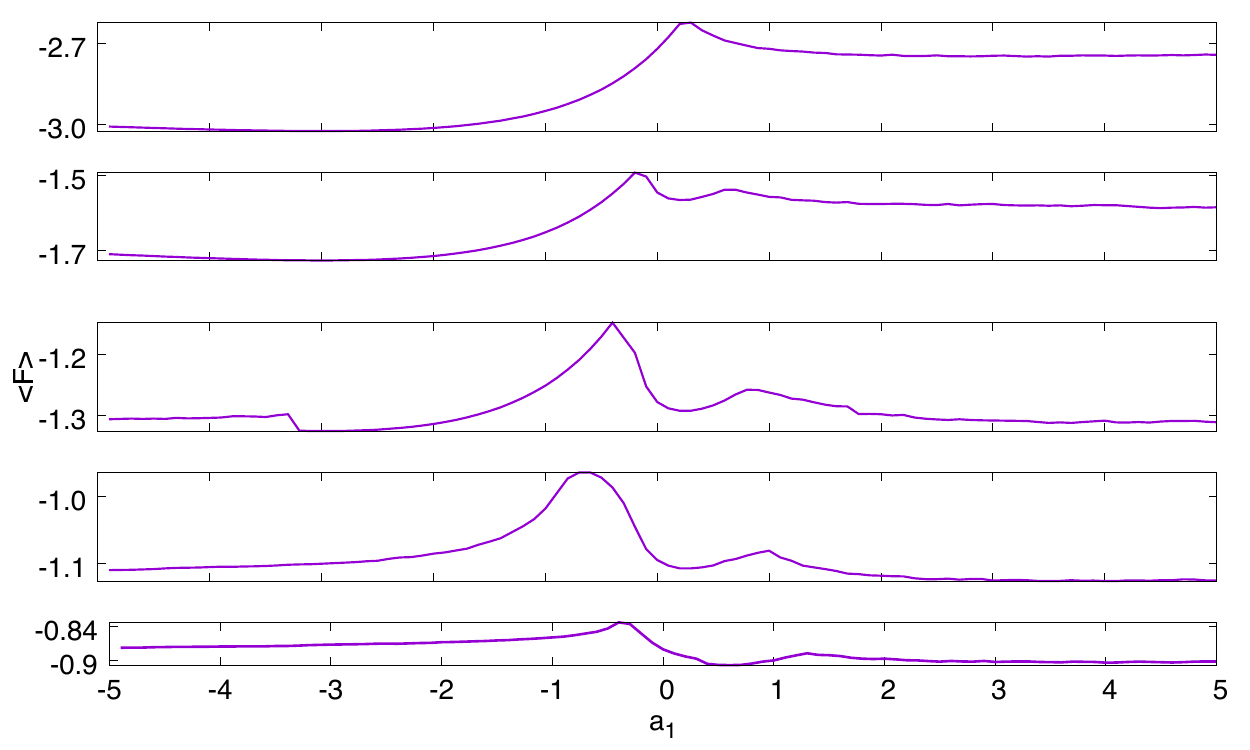} 
\caption{Expectation value of the Free energy $\langle F \rangle $ as a function of $a_1$ for different values of chemical potential from the top panel $\mu \in \{2.0,0.7,0.3,0.0,-0.4\}$, $U=4$.}
\label{fig:min_shift_U4} 
\end{figure}

In this section we are going to apply our scheme to the hole doped case for an interaction comparable to the bandwidth, namely $U=4$. The Free energy plots in Fig.~\ref{fig:min_shift_U4} indicate that the situation is more involved in this case compared to the one obtained in the strong coupling limit of Sec.~\ref{sec:holedoped_12}. There appears three local minima: one in the region $a_1\in [-4,-3]$, one in the region $a_1 \in [0,1]$, and one in the region $a_1\in[3,4]$. In our discussion below we will call these minima $m_1$, $m_2$, and $m_3$.
For $\mu>0.3$ $m_1$ is the global minimum. Upon hole doping we can see that the local minimum $m_3$ is pushed down in energy and the minimum $m_2$ gets formed. For $\mu<0.3$ the minimum in  $m_3$ becomes the global one until for $\mu<-0.3$ the minimum in $m_2$ becomes the global minimum.

\begin{figure}
\includegraphics[width=0.5\textwidth]{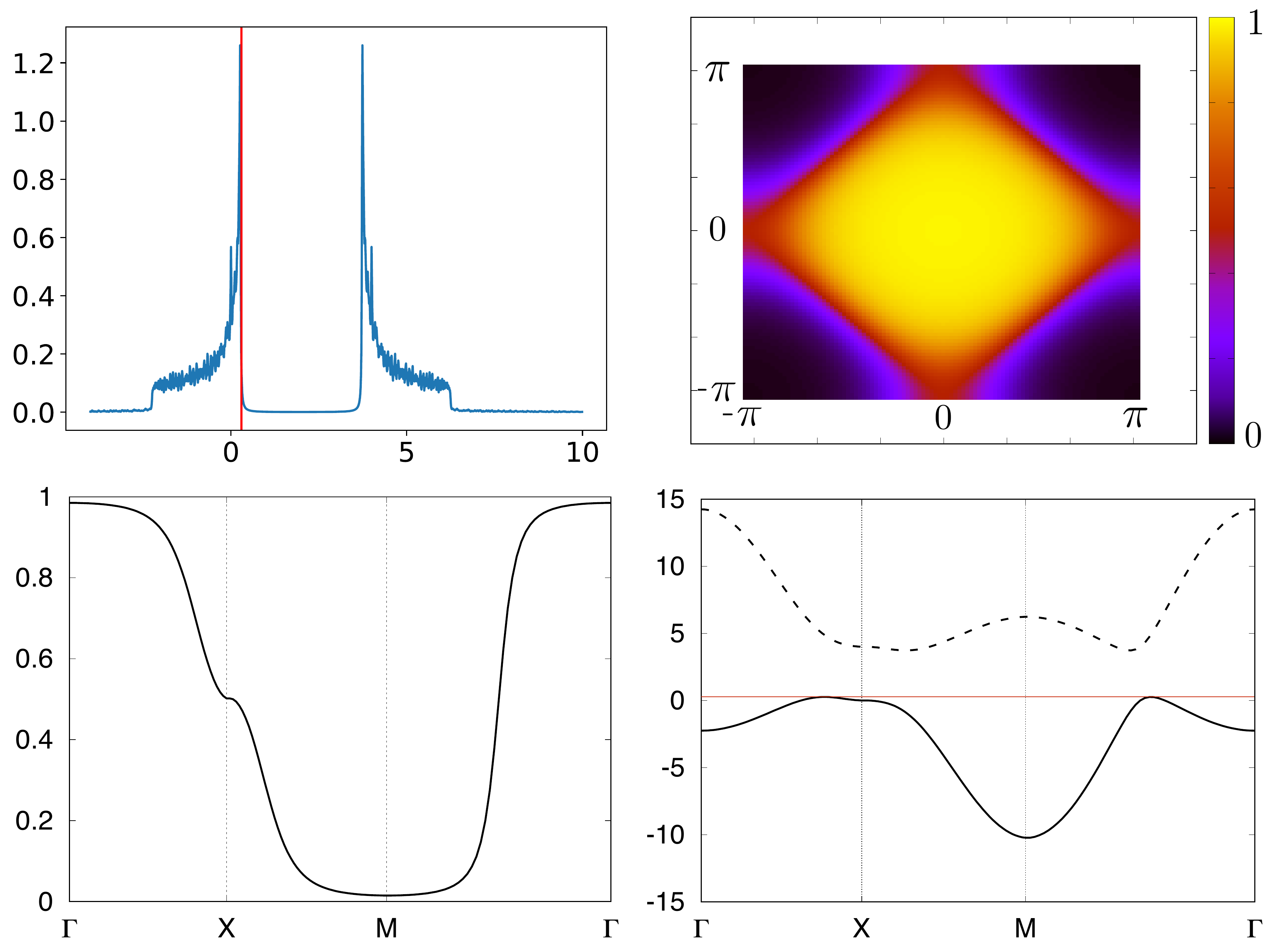} 
\caption{Characterization of solution $m_1$.
DOS (top left), colormap of the average occupation in the first Brillouin zone (top right),
average occupations along high symmetry lines (bottom left), 
and band structure (bottom right), for $U=4$, $\mu=0.3$ and $a_1 = -3.0$ and $\langle n_{\downarrow} \rangle= 0.50$.
The red line indicates the chemical potential.}
\label{fig:m1_U4} 
\end{figure}

The character of the solution $m_1$ may elucidated from Fig.~\ref{fig:m1_U4}. It is characterized by an insulating gap and predicts the system to be half filled for $\mu \in [0.3,2]$. This solution is also characterized by the absence of  a Fermi surface, and should be viewed as being in the same phase as the corresponding half-filled solution studied above with $a_1 = -3$, and also the corresponding strong coupling solution.

\begin{figure}
\includegraphics[width=0.5\textwidth]{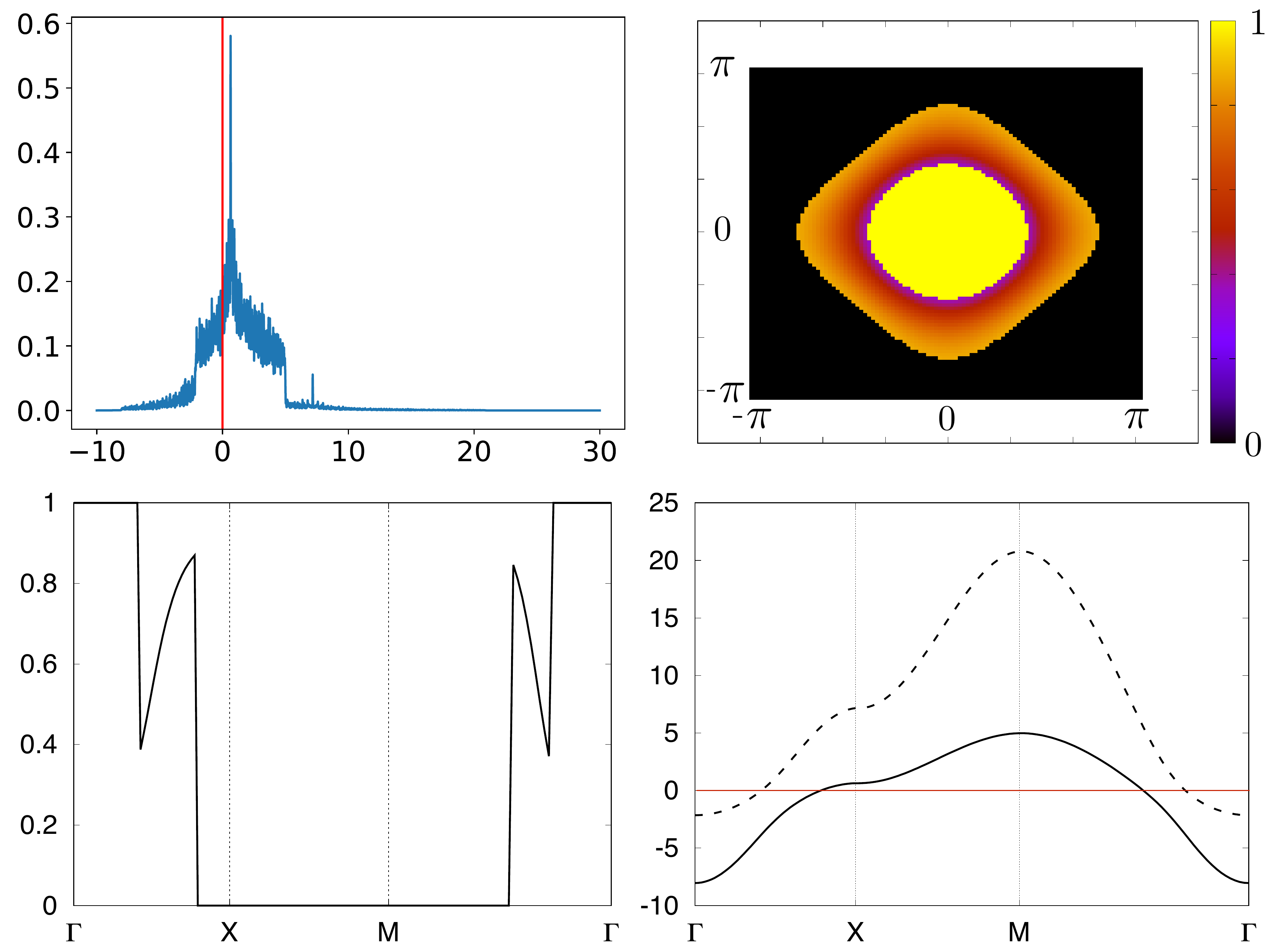} 
\caption{Characterization of solution $m_3$.
DOS (top left), colormap of the average occupation in the first Brillouin zone (top right),
average occupations along high symmetry lines (bottom left), 
and band structure (bottom right), for $U=4$, $\mu=0.0$ and $a_1=3.5$ and  $\langle n_{\downarrow} \rangle= 0.30$.
The red line indicates the chemical potential.}
\label{fig:m3_U4} 
\end{figure}

The solution $m_3$ may be characterized by studying Fig.~\ref{fig:m3_U4}, it is gapless which suggests a metallic state.
There is moreover two sharp Fermi surfaces with discontinuities in the occupation numbers and a partially depleted ``ring'' around the $\Gamma$-point is formed. This interesting state is not present in the strong interaction solution.

\begin{figure}
\includegraphics[width=0.5\textwidth]{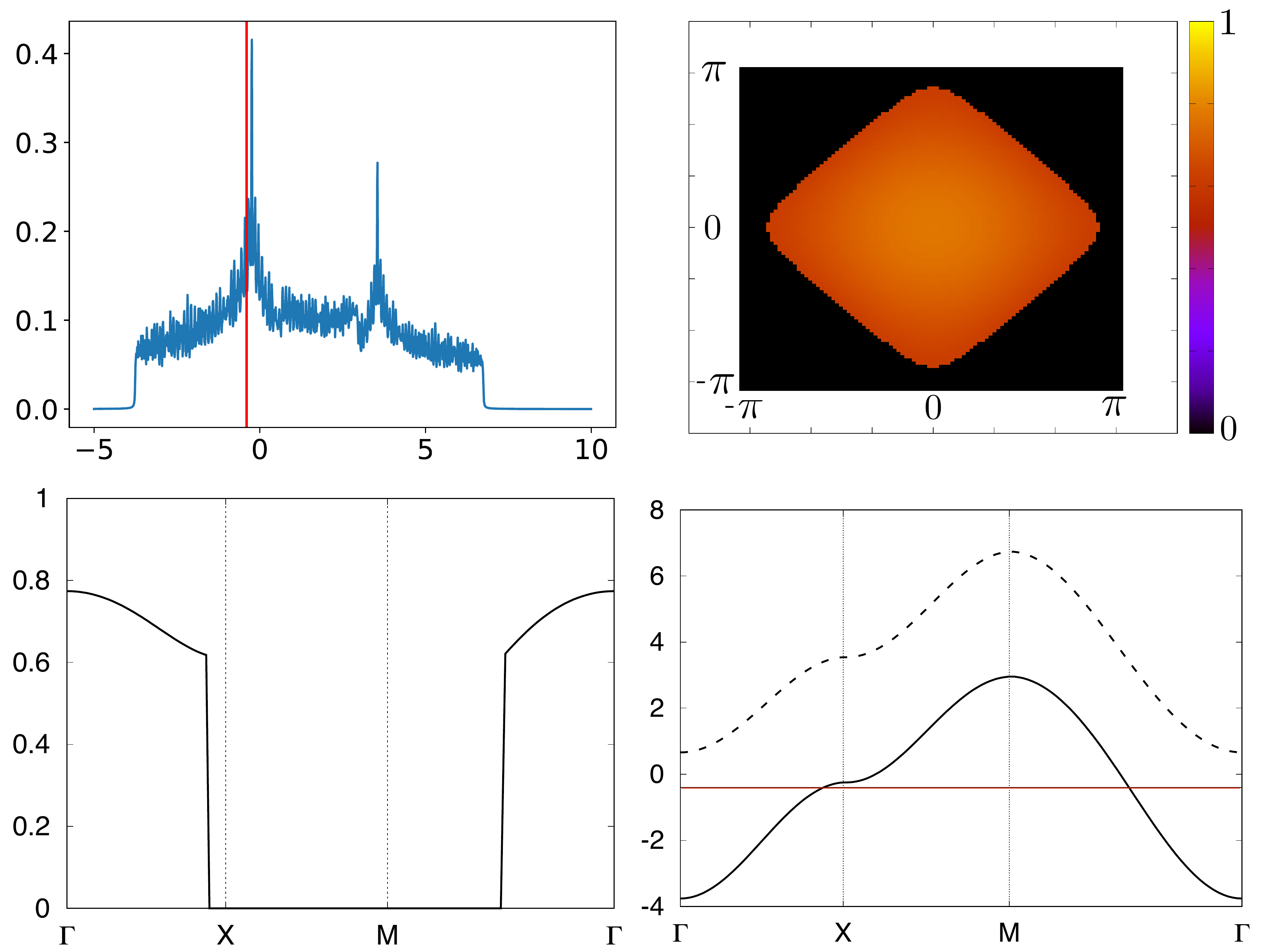} 
\caption{Characterization of solution $m_2$.
DOS (top left), colormap of the average occupation in the first Brillouin zone (top right),
average occupations along high symmetry lines (bottom left), 
and band structure (bottom right), for $U=4$, $\mu=-0.4$ and $a_1=0.6$ and $\langle n_{\downarrow} \rangle= 0.30$ .
The red line indicates the chemical potential.}
\label{fig:m2_U4} 
\end{figure}

The solution $m_2$ is also characterized by the absence of a gap as can be seen in Fig.~\ref{fig:m2_U4}. There is one sharp Fermi surface and consequently, in contrast to the solution $m_3$, there is no formation of a depleted ring around the $\Gamma$-point.

\begin{figure}
\includegraphics[width=0.5\textwidth]{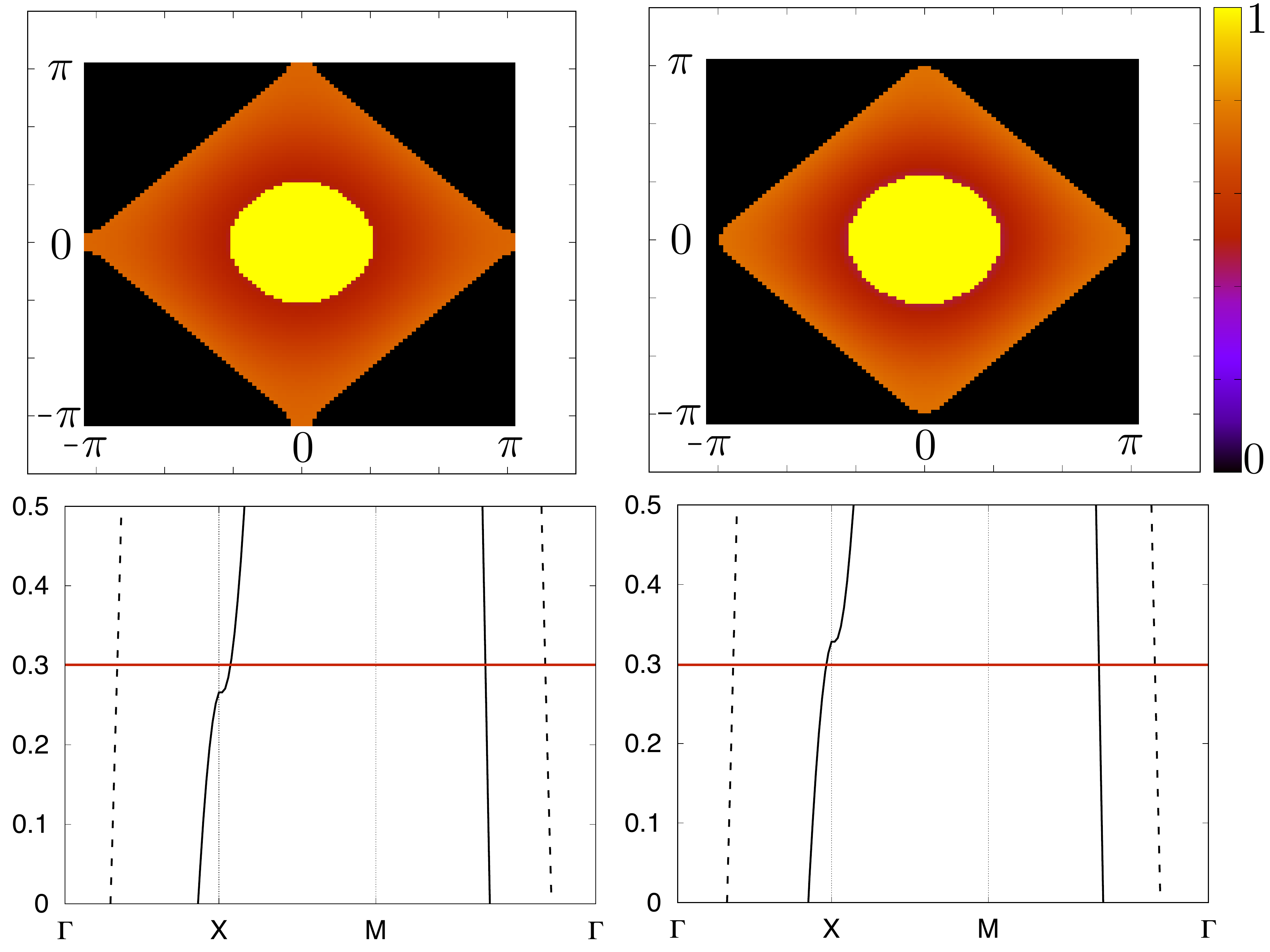}
\caption{Colormap of the occupation in the first Brillouin zone for $U=4$ and $\mu=0.3$, for $a_1=1.7$ (top left) and $a_1=1.9$ (top right). Zoom in of band structure for the corresponding two cases (lower two panels) making the topological transition in the shape of the Fermi surface apparent. The red line indicates the chemical potential.}
\label{fig:comparison_lifshitz4} 
\end{figure}

As in the strong coupling case above there exists discontinuities in some curves in Fig.~\ref{fig:min_shift_U4}. In particular for $\mu=0.3$ there is a discontinuity in $\langle F \rangle(a_1)$ around $a_1=1.8$ that is barely visible in the figure. The origin of this is a Lifshitz type transition where the Fermi surface change topology passing from a connected to a non-connected one as can be seen in Fig.~\ref{fig:comparison_lifshitz4}. The discontinuity near $a_1 \approx -3.2$ at $\mu=0.3$ is of the same type as the considered above, see Fig.~\ref{fig:comparison_lifshitz}.

\section{Conclusions and Outlook}
\label{sec:conclusions}

In the context of the Green's function equation of motion method, we disclose the dependency between the algebra of the operators and their evolution, stressing that the Hermiticity of the $E$-matrix is a fundamental relation that all physical theories must satisfy. We also realized that for an arbitrary truncation the Hermiticity of the $E$-matrix is generally violated which leads to unphysical approximation for the Green's function.

To overcome this type of problem a novel truncation scheme for the equations of motion based on a partial orthogonalization was developed, in the context of the hierarchy of the operators.
The main outcome of this procedure is an approximation for the fermionic Green's function, which can in principle be extended to an arbitrary number of poles.

We applied this truncation scheme to a two-pole approximation for the Hubbard Model showing that the Hubbard-I and Mancini results can be obtained as a particular choices of a much wider range of decoupling possibilities. 
We introduced a variational procedure to determine the partial orthogonalization parameter(s). By employing it we analyzed a set of possible solutions for the two-pole approximation for the Hubbard model and we show that independently of the choice of the orthogonalization parameter both the atomic limit and the non-interacting limit are obtained as special cases for the half-filled case. Furthermore the solutions obtained, suggests the presence of a Mott metal-insulator transition both in the large coupling limit and in the intermediate one. In the latter case we also find the presence of three competing solutions: one with an insulating character and two with metallic ones, characterized by different occupations in the first Brillouin zone and different number of Fermi surfaces.
We want to stress that the variational procedure proposed to fix the parameters in this paper is not the only option available and in principle whatever decoupling parameters which satisfy the algebra constraint should be considered valid.
Despite that, this method allows a transparent way to determine the part of the Green's function that is neglected 
from the original theory and constrain its total spectral weight.
This enables further refinements of the approximate Green's function, where the effect of the neglected part can be incorporated in the theory using an adequate form of the self-energy.

In the end it is important to recall that this scheme can be applied both in the study of fermionic and bosonic systems. Various application of this novel decoupling scheme also in case of broken symmetries are planned for future works. 

\acknowledgments
Funding from the Knut and Alice Wallenberg Foundation and the Swedish research council Vetenskapsr{\aa}det is gratefully acknowledged.
\bibliography{paper_ref.bib}

\appendix
\newpage
\begin{widetext}

\begin{table}
\centering
\caption{\label{table:zeroTE} 
Benchmark zero-temperature energies at half filling, $T=0$, for a range of interaction strengths $U$ from Ref.~\onlinecite{LeBlanc2015}  compared with ours. The benchmark results have been rounded to approximately two decimals.}
\label{table:zeroTE}
\begin{tabular}{|l|cc|cc|cc|cc|cc|} 
\hline
\multicolumn{1}{|c|}{U} & \multicolumn{2}{c|}{2}                                            & \multicolumn{2}{c|}{4}                                                 & \multicolumn{2}{c|}{6}                                                & \multicolumn{2}{c|}{8}                                                 & \multicolumn{2}{c|}{12}                                                 \\ 
\hline\hline                                      
Ref.~\onlinecite{LeBlanc2015}                & -1.17                     &                                     & -0.86                     &                                          & -0.66                     &                                         & -0.52                      &                                         & -0.37                      &                                          \\ 
\hline
2P\_PO\_-3              & \multicolumn{1}{l}{-1.2601} & \multicolumn{1}{l|}{~ ~ ~ ~ ~ ~ ~~} & \multicolumn{1}{l}{-1.0255} & \multicolumn{1}{l|}{~ ~ ~ ~ ~ ~ ~ ~ ~ ~} & \multicolumn{1}{l}{-0.8572} & \multicolumn{1}{l|}{~ ~ ~ ~ ~ ~ ~ ~ ~~} & \multicolumn{1}{l}{~-0.7313} & \multicolumn{1}{l|}{~ ~ ~ ~ ~ ~ ~ ~ ~~} & \multicolumn{1}{l}{-0.55793} & \multicolumn{1}{l|}{~ ~ ~ ~ ~ ~ ~ ~ ~~}  \\ 
\hline
2P\_PO\_1               & \multicolumn{1}{l}{-1.1459} & \multicolumn{1}{l|}{}               & \multicolumn{1}{l}{-0.7187} & \multicolumn{1}{l|}{}                    & \multicolumn{1}{l}{-0.3367} & \multicolumn{1}{l|}{}                   & \multicolumn{1}{l}{-0.0002}  & \multicolumn{1}{l|}{}                   & \multicolumn{1}{l}{0.0000}   & \multicolumn{1}{l|}{}                    \\
\hline
\end{tabular}
\end{table}

\begin{table}
\centering
\caption{\label{table:zeroTD} Benchmark zero-temperature double occupancy at half filing, for a range of interaction strengths $U$. The benchmark results have been rounded to approximately two decimals.}
\label{table:zeroTD}
\begin{tabular}{|l|cc|cc|cc|cc|cc|} 
\hline
\multicolumn{1}{|c|}{U} & \multicolumn{2}{c|}{2}                             & \multicolumn{2}{c|}{4}                             & \multicolumn{2}{c|}{6}                             & \multicolumn{2}{c|}{8}                                                            & \multicolumn{2}{c|}{12}                              \\ 
\hline\hline
Ref.~\onlinecite{LeBlanc2015}                 & 0.19                      &               & 0.13                     &        & 0.081                     &        & 0.054                                                    &      & 0.028                      &         \\ 
\hline
2P\_PO\_-3              & \multicolumn{1}{l}{0.1405} & \multicolumn{1}{l|}{} & \multicolumn{1}{l}{0.0980} & \multicolumn{1}{l|}{} & \multicolumn{1}{l}{0.0721} & \multicolumn{1}{l|}{} & \multicolumn{1}{l}{0.0548}                                & \multicolumn{1}{l|}{} & \multicolumn{1}{l}{0.03405} & \multicolumn{1}{l|}{}  \\ 
\hline
2P\_PO\_1               & \multicolumn{1}{l}{0.1540} & \multicolumn{1}{l|}{} & \multicolumn{1}{l}{0.0921} & \multicolumn{1}{l|}{} & \multicolumn{1}{l}{0.0422} & \multicolumn{1}{l|}{} & \multicolumn{1}{l}{$2\times10^{-5}$} & \multicolumn{1}{l|}{} & \multicolumn{1}{l}{0}       & \multicolumn{1}{l|}{}  \\
\hline
\end{tabular}
\end{table}

\end{widetext}

\end{document}